\documentstyle[a4,12pt,epsf,cite]{article}

\epsfverbosetrue
\newcommand{\ewxy}[2]{\setlength{\epsfxsize}{#2}\epsfbox[30 30 640 640]{#1}}

\newcommand{\kcrit}{\mbox{$\kappa_{\rm crit}$}}
\newcommand{\ksea}{\mbox{$\kappa^{\rm sea}$}}
\newcommand{\kval}{\mbox{$\kappa^{\rm val}$}}

\newcommand{\e}{{\rm e}}

\newcommand{\mut}{\multicolumn{2}{c}{\mbox{}}}
\newcommand{\mute}{\multicolumn{2}{c|}{\mbox{}}}

\def\gev{{\rm GeV}}

\def\fm{{\rm fm}}
 
\def\psibar{\overline{\psi}}
 
\def\csw{c_{\rm sw}}

\newcommand{\be}{\begin{equation}}
\newcommand{\ee}{\end{equation}}
\newcommand{\bea}{\begin{eqnarray}}
\newcommand{\eea}{\end{eqnarray}}

\newcommand{\eq}[1]{eq.\,(\ref{#1})}
\newcommand{\fig}[1]{Fig.\,\ref{#1}}
\newcommand{\tab}[1]{Table\,\ref{#1}}

\newcommand{\plus}{\makebox[15pt][c]{$+$}}
\newcommand{\minus}{\makebox[15pt][c]{$-$}}
\newcommand{\figurebox}[2]{\fbox{\vbox to#2in{\hbox to #1in{\hfil}\vfil}}}
\newcommand{\errr}[2]{\raisebox{0.08em}{\scriptsize
                                            {$\;\begin{array}{@{}l@{}}
                          \plus\makebox[0.9em][r]{#1} \\[-0.12em]
                          \minus\makebox[0.9em][r]{#2}
                        \end{array}$}}}
\newcommand{\err}[2]{\raisebox{0.08em}{\scriptsize
                          {$\;\begin{array}{@{}l@{}}
                          \plus\makebox[0.55em][r]{#1} \\[-0.12em]
                          \minus\makebox[0.55em][r]{#2}
                        \end{array}$}}}
\newcommand{\er}[2]{\raisebox{0.08em}{\scriptsize
                          {$\;\begin{array}{@{}l@{}}
                          \plus\makebox[0.15em][r]{#1} \\[-0.12em]
                          \minus\makebox[0.15em][r]{#2}
                        \end{array}$}}}
\newcommand{\texp}{\tau^{\rm exp}}
\newcommand{\tint}{\tau^{\rm int}}
\newcommand{\tcum}{\tau^{\rm cum}}
\newcommand{\bm}[1]{\mbox{\boldmath ${#1}$}}
\newcommand{\sbm}[1]{\scriptstyle\mbox{\boldmath ${#1}$}}
\newcommand{\rvec}{\bm{r}}
\newcommand{\dvec}{\bm{d}}
\newcommand{\sdvec}{\sbm{d}}
\newcommand{\kvec}{\bm{k}}
\newcommand{\gtaeq}{\raisebox{-.6ex}{$\stackrel{\textstyle{>}}{\sim}$}}

\newcommand{\Tmin}{t_{\rm min}}
\newcommand{\Tmax}{t_{\rm max}}
\newcommand{\mps}{m_{\rm PS}}
\newcommand{\mv}{m_{\rm V}}

\begin{document}

\begin{titlepage}
 
\begin{flushright}
Oxford Preprint OUTP--98--53--P\\
Edinburgh Preprint 98/11\\
Liverpool Preprint LTH 430\\
Swansea Preprint SWAT 196
\end{flushright}
 
\vspace*{3mm}
 
\begin{center}
{\Huge Light hadron spectroscopy with $O(a)$ improved dynamical
fermions}\\[12mm]
{\large\it UKQCD Collaboration}\\[3mm]

{\bf C.R.~Allton}$^1$, {\bf S.P.~Booth}$^2$, {\bf K.C.~Bowler}$^3$,
{\bf M.~Foster}$^4$, {\bf J.~Garden}$^3$, {\bf A.C.~Irving}$^4$,
{\bf R.D.~Kenway}$^3$, {\bf C.~Michael}$^4$, {\bf J.~Peisa}$^4$,
{\bf S.M.~Pickles}$^3$, {\bf J.C.~Sexton}$^5$,
{\bf Z.~Sroczynski}$^3$, {\bf M.~Talevi}$^3$,
{\bf H.~Wittig}$^6$ \\


\vspace{8mm}


$^1$\,Department of Physics, University of Wales, Singleton Park, Swansea
SA2~8PP, UK\\

$^2$ Edinburgh Parallel Computing Centre, University of
Edinburgh, Edinburgh EH9~3JZ, Scotland\\

$^3$\,Department of Physics \& Astronomy, University of Edinburgh,
Edinburgh EH9~3JZ, Scotland\\

$^4$\,Division of Theoretical Physics, Department of Mathematical
Sciences, University of Liverpool, Liverpool L69~3BX, UK\\

$^5$\,School of Mathematics, Trinity College, Dublin~2,
and
Hitachi Dublin Laboratory, Dublin~2, Ireland\\

$^6$\,Theoretical Physics, University of Oxford, Oxford OX1~3NP, UK\\

\end{center}

\vspace{3mm}
\begin{abstract}
We present the first results for the static quark potential and the
light hadron spectrum using dynamical fermions at $\beta=5.2$ using an
$O(a)$ improved Wilson fermion action together with the standard
Wilson plaquette action for the gauge part. Sea quark masses were
chosen such that the pseudoscalar-vector mass ratio, $\mps/\mv$,
varies from $0.86$ to $0.67$. Finite-size effects are studied by using
three different volumes, $8^3\cdot 24, 12^3\cdot 24$ and $16^3\cdot
24$. Comparing our results to previous ones obtained using the
quenched approximation, we find evidence for sea quark effects in
quantities like the static quark potential and the vector-pseudoscalar
hyperfine splitting.
\end{abstract}

\end{titlepage}

\section{Introduction}

Recent years have seen a lot of progress in understanding the spectrum
and the decays of hadrons using numerical simulations of lattice QCD
(for recent reviews
see\,\cite{yoshie_lat97,guesken_lat97,gottlieb_lat96}). While much
effort has been invested in controlling systematic errors such as
finite-size effects and lattice artefacts, the inclusion of quark
loops in the stochastic evaluation of the Feynman path integral still
presents a major challenge. Therefore, most simulations rely on the
quenched approximation for which the systematic errors incurred by
neglecting dynamical quark effects cannot be assessed. However, recent
progress in the development of efficient
algorithms~\cite{jansen_lat96} and increased computer power, has
greatly increased the prospects for simulations with dynamical
quarks. Several groups have published results for hadronic quantities
in the light quark sector and the static quark potential from
dynamical simulations
\cite{SESAM_pot96,SESAM_quark97,SESAM_hadron98,Columbia_dyn97,CPPACS_dyn97,
MILC_spec_dyn97,SESAM_dyn97,UKQCD_lat97}. At the same time it has been
demonstrated that leading lattice artefacts of order~$a$ in physical
observables can be eliminated through the non-perturbative
implementation of the Symanzik improvement programme
\cite{Alpha_letter,AlphaI}. The non-perturbatively $O(a)$ improved
Wilson action has been determined in the quenched approximation
\cite{AlphaIII,KlassenEdwards_lat97}, and first results have also been
reported for $n_f=2$ flavours of dynamical
quarks~\cite{JansenSommer_98}. This enables one to study the effects
of the inclusion of dynamical quarks whilst having better control over
discretisation errors.

In this work we report on calculations of the light hadron spectrum
using two flavours of $O(a)$-improved dynamical Wilson quarks at
$\beta=5.2$. For the improvement (clover) coefficient we have used
$\csw=1.76$, a preliminary estimate kindly supplied by the Alpha
Collaboration prior to the final result $\csw=2.017$ at $\beta=5.2$,
presented in \cite{JansenSommer_98}. The main aim of this study is to
understand qualitative features of dynamical simulations with improved
fermions, such as the sea quark mass dependence of observables,
finite-volume effects and estimates of autocorrelation times. The
complete removal of $O(a)$ lattice artefacts is therefore not our
highest priority in this work. Results for the hadron spectrum
obtained with the ``correct'' values of $\csw$ will be published
elsewhere \cite{UKQCD_dynamical_NP}.

The plan for the remainder of this paper is as follows. In
section~\ref{SEC_simul} we describe the details of our simulation,
including the implementation of the algorithm and the analysis of
autocorrelations. Section~\ref{SEC_potential} contains our results for
the static quark potential. The results for the light hadron spectrum
and discussions of finite-volume and sea quark effects are presented
in section~\ref{SEC_hadrons}. Section~\ref{SEC_concl} contains our
conclusions. Finally, in Appendix~\ref{APP_tables} we list our results
for hadron masses on all lattice sizes and parameter values.

\section{The simulation}
\label{SEC_simul}
In this section we fix our notation, describe the details of the
implementation of the Generalised Hybrid Monte Carlo (GHMC) algorithm
\cite{GHMC} on the Cray T3E, and give an overview of the simulation
parameters used in our calculation. We end this section with a
discussion of autocorrelations.

\subsection{Lattice action}

The lattice action can be split into a pure gauge part $S_G$ and a
fermionic part $S_F$
\be
   S[U,\psibar,\psi] = S_G[U]+S_F[U,\psibar,\psi],
\ee
where
\be
   S_G[U] = \beta\sum_P\left(1-\frac{1}{3}{\rm Re\,Tr\,}U_P\right)
\ee
is the Wilson plaquette action, and $S_F$ is defined by
\be
   S_F[U,\psibar,\psi] = S_F^W[U,\psibar,\psi] -
   \csw\frac{i\kappa}{2} \sum_{x,\mu,\nu}
   \psibar(x)\sigma_{\mu\nu}F_{\mu\nu}(x)\psi(x).
\ee
Here, $S_F^W$ is the standard Wilson action, and $\csw$ denotes the
improvement coefficient multiplying the Sheikholeslami-Wohlert term
\cite{SW85}. The bare parameters of the theory are the gauge coupling
$\beta\equiv6/g_0^2$ and the hopping parameter $\kappa$. Here we work
with a doublet of degenerate dynamical Wilson quarks, and hence gauge
configurations are characterised by the set of parameters
$(\beta,\kappa;\csw)$. For a description of the GHMC algorithm it is
useful to rewrite the fermionic part of the action in terms of a
complex, bosonic pseudofermion field $\phi$. In matrix notation we
have
\be
   S_F = \phi^\dagger\left(M^\dagger M\right)^{-1}\phi -2\sum_{\rm
   even~sites}\,\ln\det\,A_{xx}.
\ee
The odd-even preconditioned fermion matrix $M$ is given by
\be
   M_{xy} = A_{xx}-\kappa^2 D_{xz}A_{zz}^{-1}D_{zy},
\ee
where $D$ is the Wilson-Dirac matrix and $A$ denotes the matrix for
the Sheikholeslami-Wohlert term.

\subsection{Implementation of the GHMC algorithm}

The main limitation of the performance of the Cray T3E is the memory
bandwidth. The increased complexity of the memory system and the fact
that the processors support multiple instruction issue leads to a loss
of performance even for highly optimised Fortran code. We have
therefore chosen to write key routines in assembler, whilst using
Fortran\,90 for less CPU-intensive parts. Less than 25\% of the
required run time is spent executing Fortran code.

Using 32-bit instead of 64-bit precision to represent the fields
increases the speed by a factor~1.7. However, this gain has to be
weighed against the degradation of the acceptance rate, reversibility,
and the accuracy in the evaluation of the (global) energy difference
required in the Metropolis accept/reject step. The last issue has been
addressed by evaluating energy differences site by site and
performing the subsequent summation in higher precision.

The inversion of the fermion matrix was performed using the Stabilised
Biconjugate Gradient (BiCGStab) algorithm with odd-even
preconditioning \cite{BiCGStab}. The gain compared to using the
ordinary Conjugate Gradient algorithm was 40\%. Further algorithmic
improvements applied in this simulation are described in
\cite{zs_lat97}.

The version of the Cray T3E used in this work consisted of
96~processors, each capable of 900 MFlops peak speed. Using all our
algorithmic improvements and exploiting the architectural features of
the T3E we typically achieve sustained speeds of 25--30 GFlops on such
a configuration.

We now describe the integration schemes used in the molecular dynamics
part of the GHMC algorithm. As usual one introduces a set of conjugate
momenta\,$P$ for the gauge fields\,$U$. The HMC Hamiltonian is then
defined as
\be
{\cal H} = T[P] + S_G[U] + S_F[U,\phi^\dagger,\phi],
\ee
where $T[P]$ is the kinetic energy, and we have written $S_F$ in terms
of pseudofermion fields $\phi$ and $\phi^\dagger$. $T[P]$ is related
to the evolution operator ${\cal T}$ in molecular dynamics time
$\tau$, so that for any given set\,$U$ and\,$P$ of gauge fields and
conjugate momenta, respectively \cite{SexWein92}
\be
   {\cal T}=\left(\begin{array}{c}
           T_P\\ T_U
           \end{array}\right):\quad
     \left(\begin{array}{c}
           U(\tau)\\ P(\tau)
           \end{array}\right) \longrightarrow
     \left(\begin{array}{c}
           U(\tau+d\tau)\\ P(\tau+d\tau)
           \end{array}\right),
\ee
where $d\tau$ denotes a finite interval in simulation time. The
operators $T_P$ and $T_U$ are defined by
\bea
    T_P(d\tau) & : & U\longrightarrow e^{id\tau P}\,U \nonumber\\
    T_U(d\tau) & : & P\longrightarrow P-id\tau\frac{\partial}{\partial
    U}\left(S_G+S_F\right).
\eea
Since ${\cal T}$ represents a numerical integration of the equations
of motion it does not conserve ${\cal H}$ but introduces an error
$\Delta{\cal H}$. For a single application of ${\cal T}$ this error is
expected to grow as a power of the timestep $d\tau$
\cite{CreuGok89,McLachlan}
\be
   \Delta{\cal H} \propto (d\tau)^q.
\label{EQ_qscale}
\ee
Verification of this relation for a given integration scheme provides
a check for the correct implementation of the equations of motion. We
have compared three integration schemes defined by
\bea
   T_1 & = & T_P(d\tau)\,T_U(d\tau)  \\
   T_2 & = & T_P(\frac{d\tau}{2})\,T_U(d\tau)\,T_P(\frac{d\tau}{2})\\
   T_3 & = &
   T_P(\frac{a}{2}d\tau)\,T_U(ad\tau)\,T_P(\frac{a+b}{2}d\tau)\,
   T_U(bd\tau)\,T_P(\frac{a+b}{2}d\tau)\,T_U(ad\tau)\,
   T_P(\frac{a}{2}d\tau),
\eea
where $a=1/(2-2^{1/3})$ and $b=-2^{1/3}/(2-2^{1/3})$. Note that $T_2$
is the standard leapfrog integration scheme. One expects that
$T_1,\,T_2$ and $T_3$ cause $\Delta{\cal H}$ to vary as $(d\tau)^2$,
$(d\tau)^3$ and $(d\tau)^5$, respectively. This can be compared to the
values of $q$ obtained from the slope of $\ln\Delta{\cal H}$ as a
function of $\ln(d\tau)$. Such a comparison is shown in
\tab{TAB_qscale}. The numerically determined values of $q$ agree well
with the expected behaviour of $\Delta{\cal H}$ for the three
integration schemes, and thus we conclude that the integration of the
equations of motion has been implemented correctly. In our production
runs we have chosen an integration scheme which, like $T_2$, is exact
up to order $(d\tau)^3$. To this end we define the operators
\bea
    T_G(d\tau) & : & P\longrightarrow P-id\tau\frac{\partial}{\partial
    U} S_G \nonumber\\
    T_F(d\tau) & : & P\longrightarrow P-id\tau\frac{\partial}{\partial
    U} S_F,
\eea
and consider the generalised leapfrog integration scheme
\be
   T(d\tau) = T_F(\frac{d\tau}{2})\left
   [T_G(\frac{d\tau}{2n})T_P(\frac{d\tau}{n}) 
              T_G(\frac{d\tau}{2n})\right ]^nT_F(\frac{d\tau}{2}).
\ee
This particular scheme allows for a more efficient evaluation of the
derivative of $S_G+S_F$ with respect to the gauge field. It reduces to
the simple leapfrog scheme for $n=1$. In practice, we have used $n=1$ or 2.

Finally we note that although the Generalised Hybrid Monte Carlo
(GHMC) algorithm had been implemented, we have used standard Hybrid
Monte Carlo (HMC) in all production runs.

\begin{table}
\begin{center}
\begin{tabular}{ccc}
\hline
scheme & $(d\tau)^q$ & $q$ \\
\hline
$T_1$ & $(d\tau)^2$ & 1.982(4) \\
$T_2$ & $(d\tau)^3$ & 3.053(2) \\
$T_3$ & $(d\tau)^5$ & 5.056(6) \\
\hline
\end{tabular}
\end{center}
\caption{The leading variation of $\Delta{\cal H}$ for the three
integration schemes considered and measured values of $q$.}
\label{TAB_qscale}
\end{table}

\subsection{Simulation parameters}
\label{SEC_simul_param}

Our simulations have been performed at $\beta=5.2$. In order to be able
to study consistently finite-volume effects and the dependence on the
dynamical quark mass, we have used $\csw=1.76$ in all calculations
described in this paper. The standard lattice size in this simulation
was $12^3\cdot24$ and was chosen such as to guarantee a spatial volume
in physical units of more than $(1\,\fm)^3$. Smaller and larger
lattices of size $8^3\cdot24$ and $16^3\cdot24$ were used to monitor
finite-size effects.

In order to distinguish the bare quark mass used in the generation of
dynamical configurations from that used to compute quark propagators
for hadronic observables, we introduce the notation $\ksea$ to denote
the hopping parameter of the doublet of dynamical quarks, whilst
reserving $\kval$ for the valence quarks. In order to study the
dependence of observables on the sea quark mass, gauge configurations
have been generated at several values of $\ksea$. In
\tab{TAB_simpar} we list lattice sizes, the values of $\ksea$ and
$\kval$ and the number of configurations.

\begin{table}
\begin{center}
\begin{tabular}{|rcc|c|c|r@{.}lr@{.}lr@{.}lr@{.}lr@{.}l|}
\hline
\hline
$L^3\cdot T$ & $\beta$ & $\csw$ & \#conf. & $\ksea$ &
               \multicolumn{10}{c|}{$\kval$} \\
\hline\hline
$8^3\cdot24$ 
 & 5.2 & 1.76 &  78 & 0.1370 & 0&1370 & 0&1380 & 0&1390 & 0&1395 & \mute\\
 &     &      & 100 & 0.1380 & 0&1370 & 0&1380 & 0&1390 & 0&1395 & \mute\\
 &     &      & 100 & 0.1390 & 0&1370 & 0&1380 & 0&1390 & 0&1395 & \mute\\
 &     &      &  60 & 0.1395 & 0&1370 & 0&1380 & 0&1390 & 0&1395 & \mute\\
\hline
$12^3\cdot24$
 & 5.2 & 1.76 & 151 & 0.1370 & 0&1370 & 0&1380 & 0&1390 & 0&1395 & \mute\\
 &     &      & 151 & 0.1380 & 0&1370 & 0&1380 & 0&1390 & 0&1395 & \mute\\
 &     &      & 151 & 0.1390 & 0&1370 & 0&1380 & 0&1390 & 0&1395 & \mute\\
 &     &      & 121 & 0.1395 & 0&1370 & 0&1380 & 0&1390 & 0&1395 & \mute\\
 &     &      &  98 & 0.1398 & \mut   & 0&1380 & 0&1390 & 0&1395 & 0&1398\\
\hline
$16^3\cdot24$
 & 5.2 & 1.76 &  90 & 0.1390 & \mut  & \mut  & 0&1390 & 0&1395 & 0&1398\\
 &     &      & 100 & 0.1395 & \mut  & \mut  & 0&1390 & 0&1395 & 0&1398\\
 &     &      &  69 & 0.1398 & \mut  & \mut  & 0&1390 & 0&1395 & 0&1398\\
\hline
\hline
\end{tabular}
\end{center}
\caption{Summary of simulation parameters and statistics for the
computation of hadronic observables.}
\label{TAB_simpar}
\end{table}

Quark propagators were calculated for every combination of
$(\ksea,\,\kval)$ and combined to form hadronic two-point correlation
functions. In order to increase the projection onto the ground state,
the quark propagators used to form hadronic two-point functions have
been ``fuzzed'' at the source and/or sink according to the
prescription defined in \cite{fuzzing}. Statistical errors of
observables have been estimated using the bootstrap procedure
described in \cite{strange} using 250 bootstrap samples.

\subsection{Autocorrelations}

The determination of autocorrelation times is important in order to
achieve small statistical correlations among the ensemble of
configurations and to eliminate the effects of insufficient
thermalisation.

The autocovariance of an observable $\Omega$ is defined
as\,\cite{BrockDav} 
\be
  \Gamma_\Omega(t) = \left\langle\left( \Omega_s-\langle\Omega\rangle
  \right) \left( \Omega_{s+t}-\langle\Omega\rangle \right)
  \right\rangle,
\ee
where the subscripts on $\Omega$ label the the values obtained on
successive configurations. In practice, the expectation value
$\langle\ldots\rangle$ is replaced by the ensemble average over a
finite number of configurations. We define the autocorrelation
function of $\Omega$ by
\be
   \rho_\Omega(t) = \Gamma_\Omega(t)/\Gamma_\Omega(0),
\ee
and from the large-$t$ behaviour of $\rho_\Omega$ one obtains the
exponential autocorrelation time $\texp$
\be
  \rho_\Omega(t) \propto \e^{-t/\texp},\qquad t\to\infty.
\ee
The slowest mode of $\rho_\Omega$ is thus characterised by $\texp$,
which is relevant for the equilibration of the system. By contrast,
the integrated autocorrelation time $\tint_\Omega$ depends on the
observable and is required for the estimation of the statistical error
in $\Omega$, once the system is in equilibrium. It is defined by
\be
  \tint_\Omega = \frac{1}{2}\sum_{-\infty}^\infty\,\rho_\Omega(t) =
  \frac{1}{2} + \sum_{t=1}^\infty\,\rho_\Omega(t),
\label{EQ_tint}
\ee
where the latter equality holds since
$\rho_\Omega(-t)=\rho_\Omega(t)$. This definition implies that
statistically independent configurations for quantity $\Omega$ are
separated by $2\tint_\Omega$. In practice one has to truncate the
infinite sum in \eq{EQ_tint} at some finite value $t_{\rm max}$. The
resulting, so-called cumulative autocorrelation time $\tcum_\Omega$
\be
  \tcum_\Omega = \frac{1}{2} + \sum_{t=1}^{t_{\rm max}}\,\rho_\Omega(t)
\label{EQ_tcum}
\ee
is a good approximation to $\tint_\Omega$, provided that $t_{\rm max}$
has been chosen large enough so that any further increase does not
lead to an increase in $\tcum_\Omega$. In other words, a plot of
$\tcum_\Omega$ versus $t_{\rm max}$ should ideally exhibit a plateau
for large enough $t_{\rm max}$.

In order to obtain reliable estimates for $\texp$ and $\tcum_\Omega$,
autocorrelations should ideally be measured using ensembles containing
many more configurations than the value of $\tint_\Omega$. This
requirement is not easy to satisfy in simulations whose primary aim is
to compute hadronic properties, i.e. for which the calculation of
observables requires a non-negligible amount of CPU time. A convenient
quantity to determine autocorrelations is the average plaquette, which
in our simulations has been measured after every HMC update, and not
only on the subset of configurations used to compute quark
propagators. Although $\tcum_\Omega$ depends on the quantity $\Omega$,
the integrated autocorrelation times estimated from the plaquette
provide a useful guideline for the computation of hadronic
observables.

\begin{figure}[tp]
\begin{center}
\vspace{-2.0cm}
\ewxy{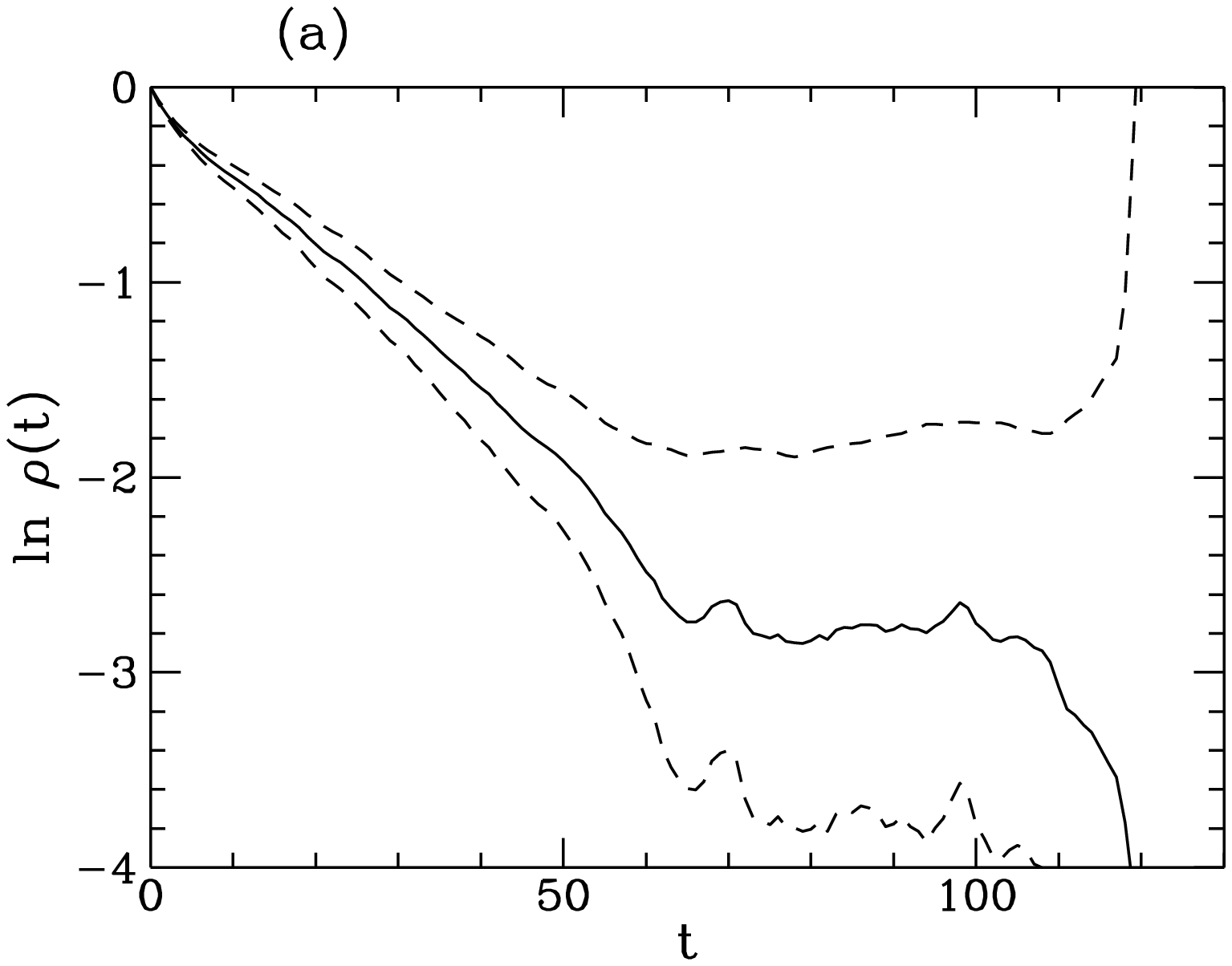}{16cm}
\vspace{-6.5cm}
\ewxy{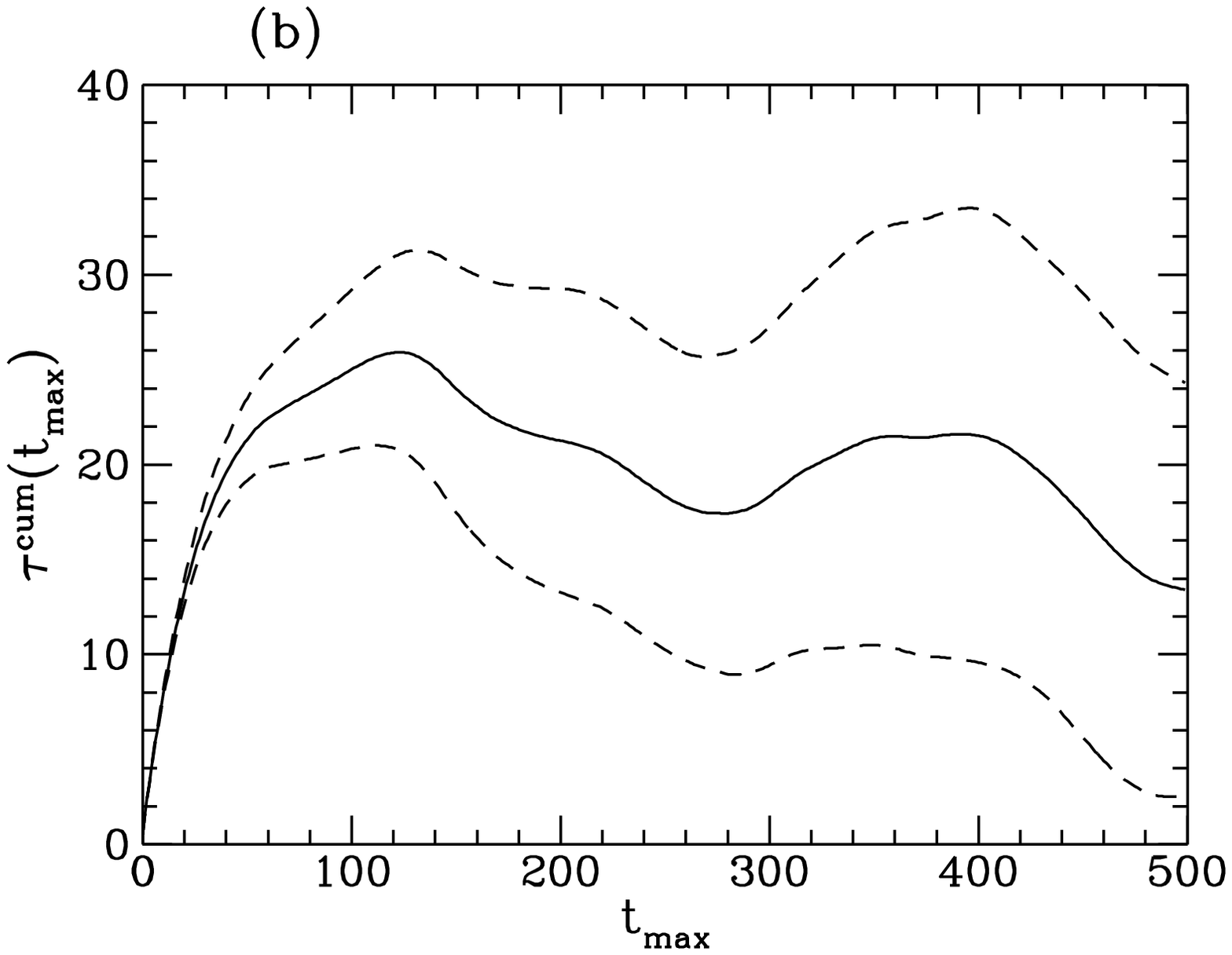}{16cm}
\end{center}
\vspace{-6.9cm}
\caption{Autocorrelations for the average plaquette on $12^3\cdot24$
and $\ksea=0.1370$. (a): $\ln\rho(t)$ plotted versus $t$; (b): the
cumulative autocorrelation time plotted against $t_{\rm max}$ (see
\protect\eq{EQ_tcum}). The solid lines follow the computed values for
$\ln\rho(t)$ and $\tcum$. The dashed lines represent the error bands
estimated from a jackknife procedure as described in the text.}
\label{FIG_autoc}
\end{figure}

Examples of our analysis of autocorrelations are shown in
\fig{FIG_autoc}. Here, the statistical errors plotted for $\rho(t)$
and $\tcum$ were estimated using a jackknife procedure. In order to
take into account the effects of autocorrelations in the error
estimate of $\rho(t)$ and $\tcum$ themselves, the original data for
these quantities were grouped in bins of size~$h$. Jackknife averages
were then formed for varying bin size, and by increasing $h$ until the
jackknife errors stabilised, the error bands in the plots were
obtained. \fig{FIG_autoc}\,(a) shows a plot of $\ln\rho(t)$ versus
$t$. $\texp$ and its error are extracted from the linear slope at
large $t$ using a fitting routine. \fig{FIG_autoc}\,(b) shows the
cumulative autocorrelation time. The central value of $\tcum$ and the
error are read off in the region where $\tcum$ shows no significant
variation within statistical errors.

Our results for $\texp$ and $\tcum$ estimated from the average
plaquette are shown in \tab{TAB_autoc}. One observes a pronounced
dependence of autocorrelation times with the mass of the sea quark. In
the range of $\ksea$ investigated in our study, $\texp$ and $\tcum$
increase by roughly a factor of two as one goes to smaller sea quark
masses. Also, a mild volume dependence of $\texp$ and $\tcum$ is
observed, so that autocorrelations appear to be slightly weaker on
larger lattices. In view of the large errors, however, this dependence
is not really significant.

\begin{table}[tp]
\begin{center}
\begin{tabular}{|rcr|cl|}
\hline
\hline
$L^3\cdot T$ & $\ksea$ & \#conf. & $\tcum$ & $\texp$ \\
\hline\hline
$8^3\cdot24$   & 0.1370 &  4900 & $>20$          & 35\err{14}{7} \\
               & 0.1380 &  6700 & 44\err{13}{10} & 43\err{3}{3}  \\
               & 0.1390 &  6600 & 36\err{10}{8}  & 53\err{6}{6}  \\
               & 0.1395 & 11800 & $>57$          & 85\err{14}{11}\\
\hline
$12^3\cdot24$  & 0.1370 &  6000 & 26\er{8}{5}    & 29\er{2}{2} \\
               & 0.1380 &  6000 & 35\err{16}{5}  & 42\er{3}{3} \\
               & 0.1390 &  5600 & 52\err{27}{25} & 43\er{5}{3} \\
               & 0.1395 &  5100 & 51\err{24}{21} & 51\er{5}{4} \\
\hline
$16^3\cdot24$  & 0.1390 &  3800 & 38\err{10}{9}  & 37\err{3}{3} \\
               & 0.1395 &  4200 & 32\err{18}{14} & 27\err{24}{9}\\
               & 0.1398 &  3000 & 32\err{23}{14} & 32\err{9}{6} \\
\hline
\hline
\end{tabular}
\end{center}
\caption{Estimates of autocorrelation times for the average plaquette
             for $\beta=5.2$, $\csw=1.76$ on several volumes.}
\label{TAB_autoc}
\end{table}

We have calculated quark propagators on configurations separated by 60
sweeps on both $8^3\cdot24$ and $12^3\cdot24$, and 40 sweeps on
$16^3\cdot24$, respectively.

\subsection{Scaling of the HMC algorithm and the value of $\kcrit$} 

Here we wish to report briefly on a simple method to obtain an
estimate of the critical value of the hopping parameter, $\kcrit$,
based on the scaling behaviour of the HMC algorithm with the quark
mass. This method is particularly useful because it can be applied
independently of an analysis of spectroscopy data, i.e. without
computing any quark propagators at all. We stress, however, that it
serves only to obtain a preliminary estimate of $\kcrit$, whose actual
value has to be extracted from the current quark mass or the quark
mass behaviour of the pseudoscalar meson.

Motivated by the idea that the computer time required for the
generation of a dynamical gauge configuration follows a scaling
behaviour near the critical quark mass, we make the ansatz
\be
  N_{\rm CG} \propto \left(\frac{1}{\kappa}-\frac{1}{\kcrit}
  \right)^\delta,
\ee
where $N_{\rm CG}$ is the number of Conjugate Gradient iterations
required to invert the fermionic part $M^\dagger M$ to some given
accuracy, and $\delta$ is a critical exponent.

\begin{figure}[tp]
\begin{center}
\vspace{-1.0cm}
\ewxy{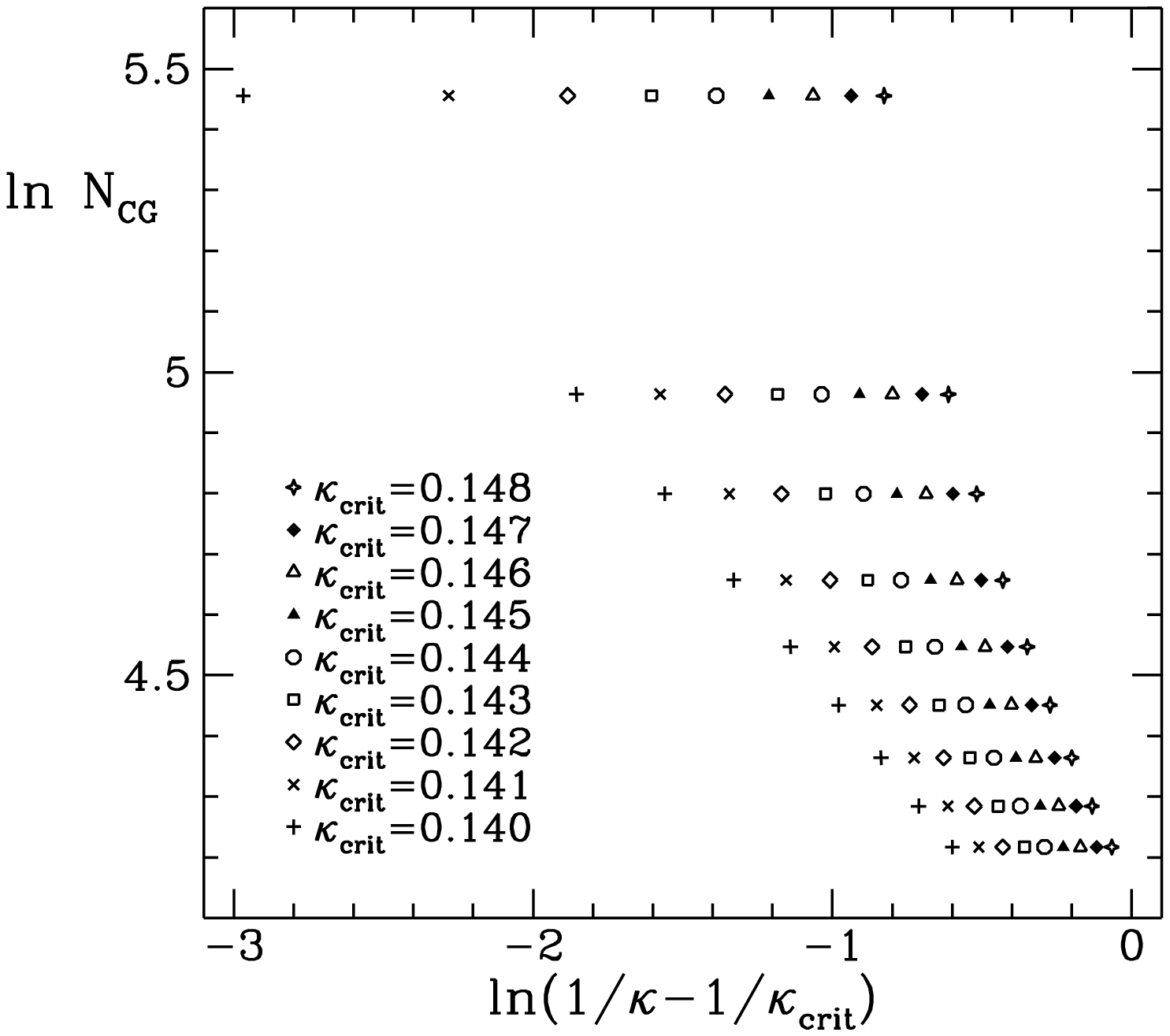}{16cm}
\vspace{-6.5cm}
\end{center}
\caption{$\ln N_{\rm CG}$ plotted versus $\ln
\left({1}/{\kappa}-{1}/{\kcrit}\right)$ for several trial values of
$\kcrit$ for $\ksea=0.136$ on $12^3\cdot24$.}
\label{FIG_logNCG}
\end{figure}

If the value of $\kcrit$ is known, we expect that $\ln N_{\rm CG}$
plotted against $\ln \left({1}/{\kappa}-{1}/{\kcrit}\right)$ should be
linear with slope $\delta$. Conversely, if $\kcrit$ is not known {\it
a priori\/}, we can use several trial values for $\kcrit$, taking the
value which reproduces the linear behaviour of $\ln N_{\rm CG}$ as the
preliminary estimate of the true $\kcrit$. Such an analysis is shown
in \fig{FIG_logNCG} for $\ksea=0.136$ on $12^3\cdot24$. Here a
straight line is obtained between $\kcrit=0.140$ and~0.141.  This
procedure can be optimised by performing a linear fit of $\ln N_{\rm
CG}$. For instance, on $8^3\cdot24$ such a fit yields
$\kcrit=0.14004(4)$. This is to be compared to the value obtained from
the quark mass behaviour of the pseudoscalar mass described in
section\,\ref{SEC_hadrons}, which gives $\kcrit=0.14047\er{6}{7}$,
which is reasonably close to the value obtained from the scaling
analysis of the HMC algorithm.

The procedure outlined in this subsection has its merits because many
inversions are performed in a typical simulation, and thus a
statistically significant value for $N_{\rm CG}$ is easily
obtained. Strictly speaking, one should only consider the first
inversion of the computation of a new trajectory, since this is the
only one guaranteed to be performed on a physical configuration
(i.e. immediately after the global accept/reject step).

\section{The static quark potential}
\label{SEC_potential}
In this section we describe the computation of the static quark
potential using our dynamical configurations. The force between static
quarks, calculated from the potential serves to determine the lattice
scale using the hadronic radius $r_0$\,\cite{sommer_r0}. Furthermore,
we study finite-size effects and investigate possible evidence for
string breaking.

\subsection{General procedure}

The method to extract the potential $V(\rvec)$ from Wilson loops
$W(\rvec,t)$ of area $|\rvec|\times t$ is standard. We have used the
algorithm described in\,\cite{albanese87} to compute ``fuzzed'' gauge
links with a link/staple weighting of 2:1 and between 10 and 20
iterations in the fuzzing algorithm. Using two different fuzzing
levels, we have constructed a $2\times2$ variational basis of Wilson
loops\,\cite{var_basis} and subsequently determined the
eigenvalues and -vectors of the generalised eigenvalue
equation\,\cite{EVeq,LW90}
\be
  W_{ij}(\rvec,t)\,\phi(\rvec)_j^{(k)} = \lambda^{(k)}(\rvec;t,t_0)\,
  W_{ij}(\rvec,t_0)\,\phi(\rvec)_j^{(k)},\quad i,j,k=1,2.
\ee
The eigenvector $\phi(\rvec)^{(1)}$, corresponding to
$\lambda^{(1)}(\rvec;t,t_0)$ at $t=1,\,t_0=0$ was then used to project
onto the approximate ground state\,\cite{sommer_r0,how_lat94}. This
combination of $(t,t_0)$ turned out to be a compromise between good
projection properties and the need to avoid the introduction of
additional statistical noise. The resulting correlator was then fitted
to both single and double exponentials for timeslices up to $t=8$. As
a cross-check we also performed exponential fits to the full
$2\times2$ matrix correlator. No significant deviations in the fit
parameters as a result of different fitting procedures have been
observed.

\subsection{Determination of $r_0/a$ on different volumes}

The computed values of the potential $V(\rvec)$ can be used to
determine the force $F(r)$ between a static quark-antiquark pair
separated by a distance $r=|\rvec|$. As discussed
in\,\cite{sommer_r0}, the force can be matched at a characteristic
scale $r_0$ to phenomenological potential models describing
quarkonia. More precisely, $r_0$ is defined through the relation
\be
   F(r_0)r_0^2 = 1.65,
\label{EQ_r0def}
\ee
which corresponds to $r_0\simeq0.49\,\fm$. Eq.\,(\ref{EQ_r0def}) can
thus be used together with lattice data for the force to extract
$r_0/a$, which then yields a value for the lattice scale in physical
units. This definition has the advantage that one needs to know the
force only at intermediate distances. An extrapolation of the force to
infinite separation, which is conventionally performed to extract the
string tension, is thus avoided. Hence, the procedure is well-suited
to the case of full QCD, for which the concept of a string tension as
the limiting value of the force appears rather dubious, because the
string is expected to break at some characteristic distance~$r_b$.

Our determination of $r_0/a$ follows closely the procedures described
in\,\cite{sommer_r0,how_lat94}. We have computed the force
$F_{\sdvec}(r_I)$ for orientations~$\dvec$ of Wilson loops according
to
\bea
  F_{\sdvec}(r_I) &=& |\dvec|^{-1}\big(V(\rvec)-V(\rvec-\dvec)\big), \\
  r_I^2 &=& -|\dvec|\Big( G_L(\rvec) - G_L(\rvec-\dvec) \Big)^{-1},
\eea
where $G_L(\rvec)$ is the lattice Greens function for one-gluon
exchange
\be
  G_L(\rvec) = 4\pi\int_{-\pi}^{\pi}\frac{d^3k}{(2\pi)^3}\,
  \frac{\cos(\kvec\cdot\rvec)}{4\sum_{j=1}^3\sin(k_j/2)}.
\ee
This definition ensures that $F(r_I)$ is a tree-level improved
quantity\,\cite{sommer_r0}. In our study we have concentrated on
``on-axis'' orientations of Wilson loops, i.e. where $\dvec=(1,0,0)$.
We have obtained estimates of $r_0$ in lattice units by a local
interpolation of $F(r_I)r_I^2$ to the point defined in
\eq{EQ_r0def}. We emphasise that this procedure does not rely on any
model assumptions about the $r$-dependence of the force.

Systematic errors in $r_0/a$ were estimated through variations in the
interpolation step (e.g. by considering a third data point for
$F(r_I)r_I^2$ besides those which straddle 1.65), and also by using
alternative fitting procedures in the extraction of the potential
(e.g. single or double exponential fits, different fitting
intervals). We note that the systematic error in $r_0/a$, in
particular for smaller values of the sea quark mass, is dominated by
the uncertainty incurred by considering different points in the
interpolation step. A summary of our results on all lattices and for
all values of $\ksea$ is shown in \tab{TAB_r0res}. The configurations
on which the potential has been determined were separated by 40 HMC
trajectories for all lattice sizes and quark masses.

\begin{table}
\begin{center}
\begin{tabular}{|rrcr@{.}lr@{.}l|}
\hline
\hline
$L^3\cdot T$  & \#conf. & $\ksea$ &
\multicolumn{2}{c}{$r_0/a$} & \multicolumn{2}{c|}{$a\,[\fm]$} \\
\hline\hline
 $8^3\cdot24$ & 119 &   0.1370 & 2&236\err{39}{46}\err{59}{12} &
 0&2192\err{46}{38}\err{11}{57} \\
              &  72 &   0.1380 & 2&475\err{74}{67}\err{54}{23} &
 0&1980\err{53}{57}\err{18}{43} \\
              &  75 &   0.1390 & 2&891\err{60}{58}\errr{156}{124} &
 0&1695\err{34}{35}\err{75}{87} \\
              & 125 &   0.1395 & 3&718\err{89}{72}\err{75}{44} &
 0&1318\err{26}{31}\err{15}{26} \\
\hline
$12^3\cdot24$ & 123 &   0.1370 & 2&294\err{20}{23}\er{7}{2} &
 0&2136\err{22}{19}\er{2}{7} \\
              & 110 &   0.1380 & 2&568\err{34}{30}\err{34}{88} &
 0&1908\err{22}{25}\err{67}{25} \\
              & 100 &   0.1390 & 3&046\err{58}{52}\err{42}{7} &
 0&1609\err{28}{30}\err{3}{22} \\
              & 103 &   0.1395 & 3&435\err{48}{47}\err{42}{0} &
 0&1426\err{19}{20}\err{0}{17} \\
              & 100 &   0.1398 & 3&652\err{29}{25}\err{7}{13} &
 0&1342\err{9}{11}\err{6}{50} \\
\hline
$16^3\cdot24$ & 100 &   0.1390 & 3&026\err{32}{24}\err{16}{0} &
 0&1619\err{13}{17}\er{0}{8} \\
              &  90 &   0.1395 & 3&444\err{40}{57}\err{26}{78} &
 0&1423\err{24}{16}\err{30}{11} \\
              &  79 &   0.1398 & 3&651\err{31}{30}\err{14}{12} &
 0&1342\err{11}{12}\er{5}{5} \\
\hline
\hline
\end{tabular}
\end{center}
\caption{Results for $r_0/a$ for different lattice sizes and quark
 masses. The lattice spacing was obtained using $r_0=0.49\,\fm$. The
 first error is the statistical, the second an estimate of the
 systematic error as described in the text.}
\label{TAB_r0res}
\end{table}

\begin{figure}[tp]
\begin{center}
\vspace{-2.0cm}
\ewxy{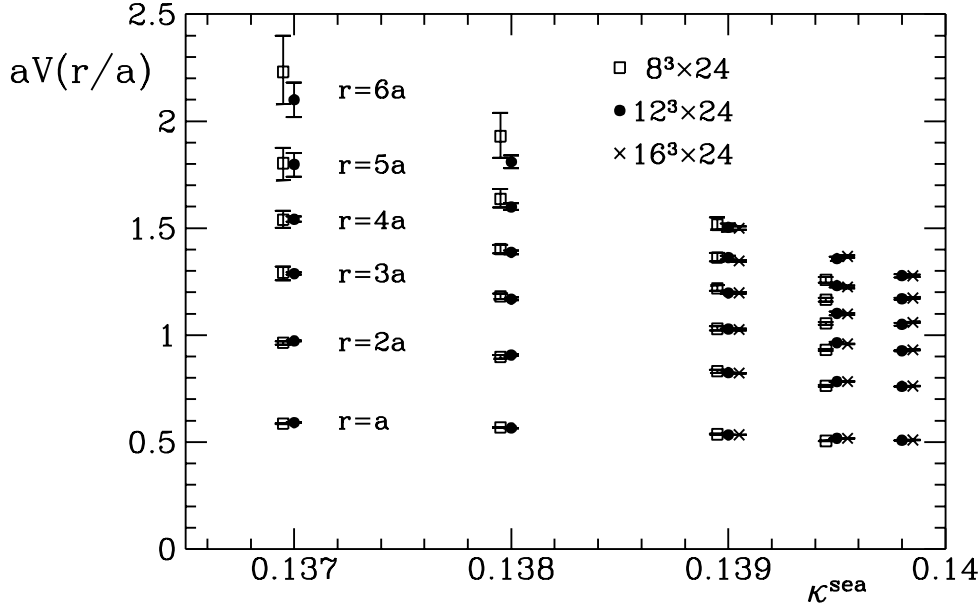}{16cm}
\end{center}
\vspace{-6.9cm}
\caption{The static quark potential for the first six on-axis
 separations $r/a=1,\ldots,6$ plotted against $\ksea$ for different
 lattice sizes.} 
\label{FIG_Vvsk}
\end{figure}

\begin{figure}[tp]
\begin{center}
\vspace{-2.0cm}
\ewxy{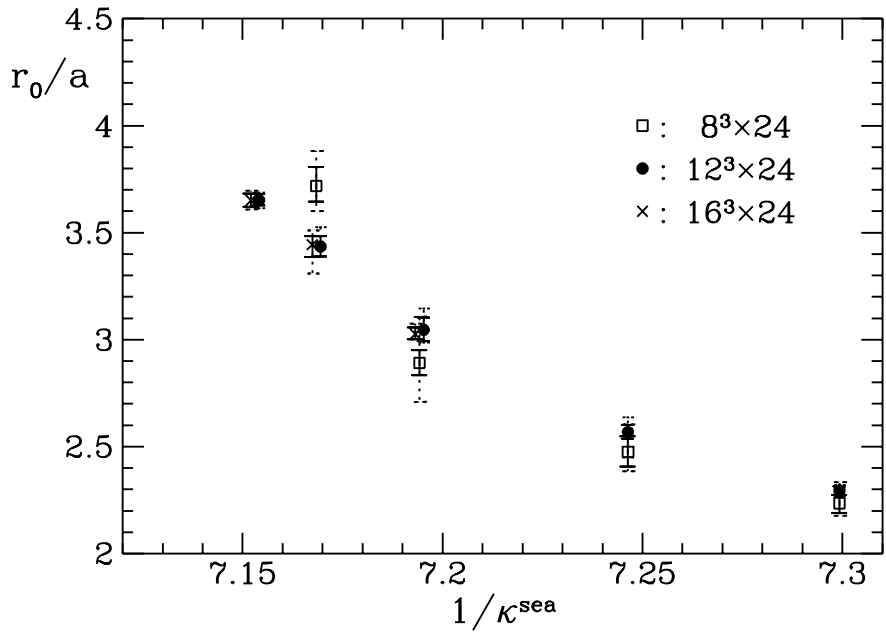}{16cm}
\end{center}
\vspace{-6.9cm}
\caption{The hadronic scale $r_0/a$ plotted against $1/\ksea$ for
 different lattice sizes. The chiral limit is approximately at the
 left margin of the figure. Solid error bars represent the statistical
 errors, whereas the dotted error bars denote the linearly added
 systematic error.}
\label{FIG_r0vsk}
\end{figure}

The comparison of results obtained on the $8^3\cdot24$ and
$12^3\cdot24$ lattices shows that there are pronounced finite-size
effects at $\ksea=0.1395$, whereas for larger quark masses these
effects are small. The presence, respectively absence, of finite-size
effects is easily recognised in the values of the potential $V(\rvec)$
itself, as shown in \fig{FIG_Vvsk}. On the other hand, there is
remarkable agreement in the data obtained on $16^3\cdot24$ and
$12^3\cdot24$, even at the lightest quark mass considered. This is
illustrated in \fig{FIG_r0vsk}, where the results of $r_0/a$ on all
lattices are plotted against $1/\ksea$. Our findings can be translated
into a bound on $L/r_0$, above which finite-size effects in the static
quark potential are largely absent at this level of precision. From
our results we infer that the bound is
\be
  L/r_0 ~\gtaeq~ 3.2,
\label{EQ_bound}
\ee
which, for instance, is still satisfied for $\ksea\leq0.1380$ on
$8^3\cdot24$. This bound, however, should not be generalised {\em
prima facie\/} to other quantities, in particular the spectrum of
hadronic states discussed in section\,\ref{SEC_hadrons}, for which
finite-volume effects could well be different. We will return to this
point in subsection\,\ref{SEC_had_finvol}.

Fig.\,\ref{FIG_r0vsk} shows that the data for $r_0/a$ obtained at the
three lightest quark masses on $L/a=12$ and~16 show a linear
behaviour. We have therefore attempted a linear extrapolation of
$r_0/a$ to the chiral limit using the data at $\ksea=0.1390, 0.1395$
and 0.1398 only, despite the lack of a theoretical motivation as to
why such an ansatz for the quark mass dependence of $r_0/a$ should be
valid. Taking into account only statistical errors in $r_0/a$, the
extrapolations for $L/a=12, 16$ yield
\bea
  L/a=12:&\quad& r_0/a=4.10\pm0.06,\quad a=0.122\pm0.002\,\fm \\
  L/a=16:&\quad& r_0/a=4.15\pm0.06,\quad a=0.121\pm0.002\,\fm.
\eea
Thus, the overall box sizes in the chiral limit amount to
$1.46(2)\,\fm$ for $L/a=12$ and $1.93(3)\,\fm$ for $L/a=16$. A
comparison with data for $r_0/a$ obtained in the quenched approximation
shows that our estimates in the chiral limit for $n_f=2$ massless
quarks at $\beta=5.2$, $\csw=1.76$ roughly correspond to values
around $\beta=5.85-5.9$ in quenched QCD\,\cite{SCRI_pot,ALPHA_pot}.

\subsection{Is there evidence for dynamical quark effects?}

\begin{figure}[tp]
\begin{center}
\vspace{-0.5cm}
\ewxy{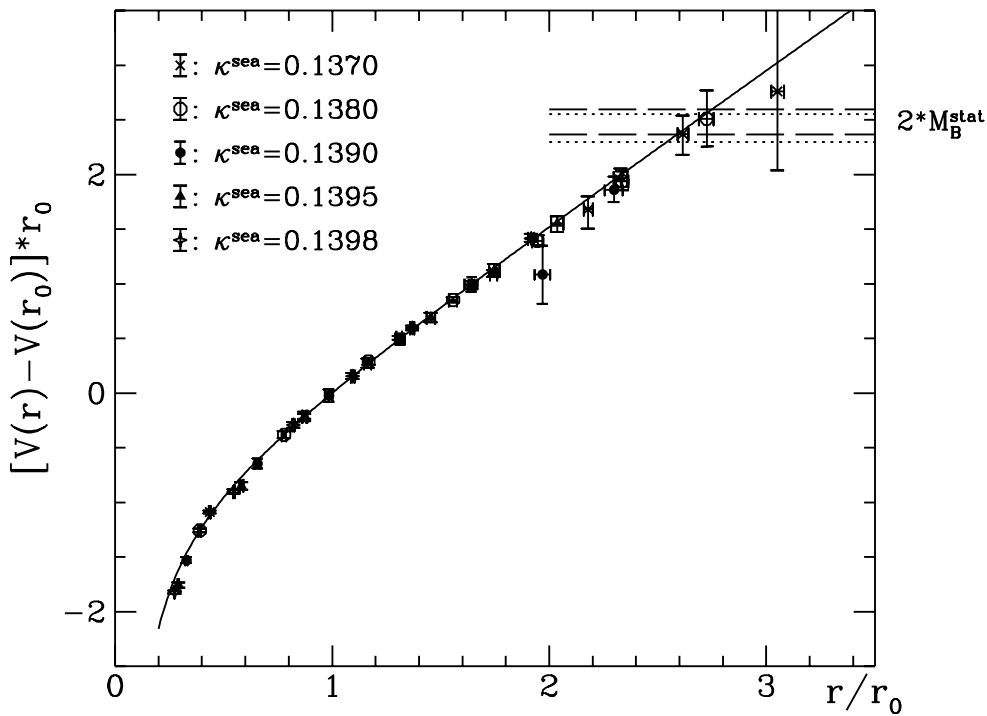}{16cm}
\end{center}
\vspace{-6.9cm}
\caption{The scaled and normalised potential as a function of $r/r_0$
as obtained on $12^3\cdot24$. The solid curve is the prediction from
\protect\eq{EQ_potstring}.} 
\label{FIG_pot}
\end{figure}

We now examine our data for the static quark potential for possible
evidence for the effects of dynamical quarks. In \fig{FIG_pot} we show
the potential in units of $r_0$, normalised to $V(r_0)$, for five
values of $\ksea$ used on $12^3\cdot24$. We compare our results to the
expression
\be
  \big[V(r)-V(r_0)\big]r_0 = (1.65-e)\left(\frac{r}{r_0}-1\right)
  -e\left(\frac{r_0}{r}-1\right), 
\label{EQ_potstring}
\ee
which follows from the standard linear-plus-Coulomb ansatz for $V(r)$,
viz.
\be
   V(r) = V_0 + \sigma r-\frac{e}{r}
\ee
and the condition \eq{EQ_r0def}. Here, $\sigma$ denotes the string
tension, and we have set $e=\pi/12$ \cite{ML_string81}, so that the
solid line in the figure has not been obtained through a fit. The data
at different $\ksea$ have been offset by $V(r_0)$, whose value was
obtained by a local interpolation of the potential.

In the presence of dynamical quarks one expects a deviation of the
data at large separations from the linear behaviour described by the
curve in \eq{EQ_potstring}, so that the potential in full QCD flattens
out due to string breaking. For separations larger than the breaking
scale $r_{\rm b}$, one expects that the potential is equal to the mass
of two ``mesons'', consisting of a static quark and a light antiquark,
i.e. the energy of a state corresponding to a broken string. The
masses of such static-light mesons have been calculated on
$12^3\cdot24$ for $\ksea=0.1390$ and~0.1395 using the technique
described in \cite{cmi_janne_98}. The error bands of this
determination are shown as the dotted ($\ksea=0.1395$) and dashed
($\ksea=0.1390$) lines in \fig{FIG_pot}.

Our data for the potential for distances $r>1\,\fm$ are neither in
disagreement with the curve in \eq{EQ_potstring}, nor with the
expected asymptotic value of $2M_B^{\rm static}$. In order to check
whether more statistics could help in revealing the flattening of the
potential at large distances, we have computed the potential on 194
stored HMC trajectories for $\ksea=0.1395$ on our larger lattice size
of $16^3\cdot24$, but no qualitative change compared to the data in
\fig{FIG_pot} could be detected. Thus, there is at present no
conclusive evidence for string breaking at length scales up to
$r\simeq1.5\,\fm$. There are a number of arguments why this is
so. Firstly, we wish to stress that the data points which probe the
largest separations in \fig{FIG_pot} have typically been obtained
using smaller values of $\ksea$, for which the sea quarks may still be
too heavy in order to produce a significantly different qualitative
behaviour of $V(r)$. It has also been argued\,\cite{guesken_lat97}
that the Wilson loop used to extract $V(r)$ does not project well onto
states of broken strings. This has, in fact, been confirmed in
simulations using bosonic matter
fields~\cite{cmi_lat91,string_higgs}. In QCD, further investigations
of this issue are required, in particular for smaller sea quark
masses. Without a clear demonstration of string breaking, we can only
give a rough estimate for the breaking scale $r_b$ from the
intersection of the data of the potential with the value $2\,M_B^{\rm
static}$. From \fig{FIG_pot} we read off $r_b\simeq2.6\,r_0$.

At small distances, where the potential is dominated by the Coulombic
part, we find that the expression in \eq{EQ_potstring} still describes
the data surprisingly well, although the points obtained for the three
lightest quark masses have a tendency to lie somewhat below the curve
at the smallest separations. This implies that the data at small
distances and quark masses seem to favour a larger value for $e$
compared to $e=\pi/12$ in the pure gauge theory, as has been
observed also in \cite{SESAM_pot96}. This qualitative observation is
consistent with the expected influence of dynamical quarks on the
short-distance regime of the potential through the $n_f$-dependence in
the running of the strong coupling constant.

However, if one wants to quantify the change in $e$, one should take
into account the lattice Greens function $G_L(\rvec)$ for one-gluon
exchange in order to account for lattice artefacts at small
distances. If $r_0/r$ in the last term of \eq{EQ_potstring} is
replaced by $r_0\,G_L(\rvec)$, it turns out that the deviation between
the data points and the curve \eq{EQ_potstring} for the two smallest
values of $r/r_0$ is smaller but still significant. We estimate that
the enhancement in $e$ amounts to about 5--10\% at the smallest value
of the sea quark mass.

This is only a crude analysis of sea quark effects in the
short-distance part of the potential. In principle these effects on
the running coupling could be probed by computing the coupling
constant from the force according to
\be
   \alpha(r) \equiv \frac{3}{4}F(r)r^2,
\ee
and comparing its scale dependence to the two-loop perturbative
$\beta$-function for $n_f=2$. In view of the many caveats concerning
our present data, such as the fairly large length scales, the
relatively heavy sea quarks and the lack of a continuum extrapolation,
we have not seriously attempted such an analysis at this stage.

To summarise, as far as the issue of string breaking is concerned, we
find no hard evidence for the effects of dynamical quarks for
distances up to 1.5\,\fm\ in the static quark potential. However,
there are indications of a qualitatively different behaviour in the
Coulombic range at small distances, which is hard to quantify and
corroborate with the present data.

\section{Hadron spectroscopy}
\label{SEC_hadrons}

In this section we describe the computation of the light hadron
spectrum. The simulation parameters used have been discussed in
subsect.~\ref{SEC_simul_param}.

\subsection{Analysis and fitting procedure}

\begin{figure}[t]
\begin{center}
\vspace{-3.5cm}
\ewxy{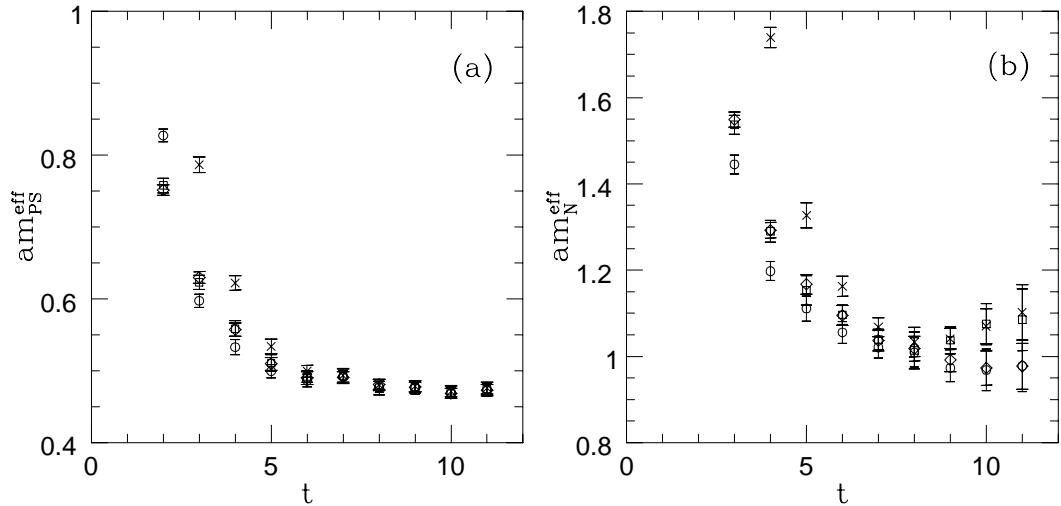}{16cm}
\end{center}
\vspace{-6.0cm}
\caption{Effective mass plots on a $12^3\cdot 24$ volume
for $\ksea=\kval=0.1398$. In (a) the pseudoscalar and in (b) the
nucleon is shown for different fuzzing combinations, i.e. FF~($\circ$),
FL~($\Diamond$), LF~($\Box$) and LL~($\times$).}
\label{FIG_fuzzing}
\end{figure}

The amplitudes and masses of hadrons are obtained in a standard way by
correlated least-$\chi^2$ fits of the correlation functions. The
fitting function used was different for mesonic and baryonic channels.
In the mesonic case, we have taken into account the backward
propagating particle on a periodic lattice by fitting to the function
\begin{equation}
C_M(t)=A_0(e^{-m_0t}+e^{m_0(T-t)})+A_1(e^{-m_1t}+e^{m_1(T-t)}), 
\label{EQ_fitmeson}
\end{equation}
where $A_i,m_i,\ i=0,1$ are the amplitudes and masses of the ground
and first excited states. In the baryonic case, we have used
\begin{equation}
C_B(t)=A_0e^{-m_0t}+A_1e^{-m_1t}.
\label{EQ_fitbaryon}
\end{equation}
We have computed hadronic two-point correlation functions with
different combinations of ``fuzzing''~\cite{fuzzing} both at source
and at sink: we denote as FF the correlator fuzzed at source and sink,
FL fuzzed only at source, etc.  We have found that the FF correlator
allows the fastest isolation of the fundamental state, even in the
case of the lightest $\kappa$'s in which the effect of fuzzing is most
important. To illustrate this point, we show in Fig.~\ref{FIG_fuzzing}
the effective mass plots of the pseudoscalar and the nucleon for
$\ksea=\kval=0.1398$ on a $12^3\cdot 24$ volume. We conclude that the
effect of fuzzing is quite significant compared with the LL case, as
also found in the quenched approximation \cite{UKQCD_quenched_NP}.  We
have fitted simultaneously the LL and FF correlators to a double
exponential functional form, using the difference between the
correlators to control the first excited state.

The quantity $\chi^2/$d.o.f.\ used to monitor the quality of a
correlated fit is known to suffer from a systematic bias, which
depends on the d.o.f.\ and the statistics \cite{Michael}. We have
implemented the technique of eigenvalue smoothing \cite{Michael} in
the computation of the correlation matrix to take this bias into
account. We have performed a ``sliding window'' analysis, fixing the
maximum value of the fit interval, i.e.~$\Tmax=11$, and varying
$\Tmin$ to monitor the optimal interval. The stability criterion we
have used is that the hadron mass should not change appreciably as
$\Tmin$ is changed by one unit. The value of $\Tmin$ has been
determined for each different combination of $\ksea$ and $\kval$.

\subsection{The dynamical spectrum}

In a numerical simulation with dynamical fermions the parameters
$\ksea$ and $\kval$ are distinct and each set of configurations
generated for different $\ksea$ is independent. We have performed the
analysis of the hadron spectrum for each fixed value of $\ksea$.  At
an intermediate level, these simulations can be thought of as
``pseudoquenched'', which come closer to the description of the real
world as the sea quark mass approaches its physical value. There is
another reason why we found it useful to simulate different values of
$\kval$ for each $\ksea$, cf.~Table~\ref{TAB_simpar}. We interpret the
heavier $\kappa$'s as describing the valence quarks, in particular the
strange, in the sea of light quarks, i.e. the up and the down.

\begin{figure}[tp]
\begin{center}
\vspace{5mm}
\ewxy{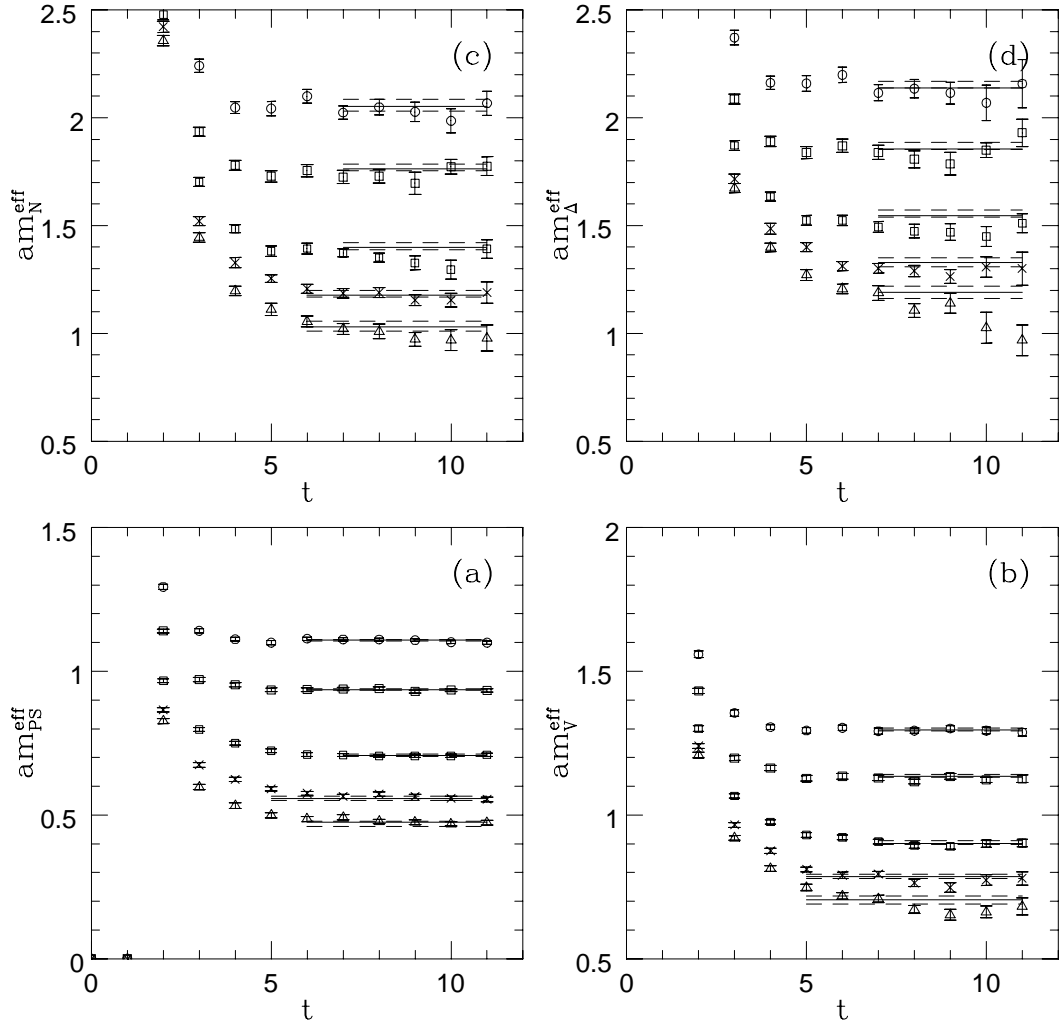}{16cm}
\end{center}
\vspace{-5.5cm}
\caption{Effective mass plots for the pseudoscalar~(a), vector~(b),
nucleon~(c) and $\Delta$~(d), on a $12^3\cdot 24$ volume for all
values of $\ksea$. The effective mass has been calculated from the FF
correlator with $\ksea=\kval$. Solid lines denote the fitted value of
the mass, obtained from a simultaneous fit to the FF and LL
correlators. The dashed lines denote the error band, and the length of
the lines indicates the fit interval.}
\label{FIG_effmass}
\end{figure}

In Fig.~\ref{FIG_effmass} we show the effective mass plots on a
$12^3\cdot 24$ volume for the pseudoscalar, vector, nucleon and
$\Delta$, obtained for $\ksea=\kval$. The pseudoscalar shows a clear
plateau at all values of the hopping parameter, whereas for the vector
the plateau becomes more unstable at the lightest quark masses. In the
baryonic channels, on the other hand, the plateaux are, as expected,
more fluctuating and we require a longer lattice in the time
direction.

In the tables given in Appendix~\ref{APP_tables} we summarise the
results for the hadron masses obtained on the different volumes. We
give the masses both in lattice units and in units of $r_0$, the
latter being more significant in the comparison between different
volumes, as it compensates for the sea quark dependence of the lattice
spacing.

In \tab{TAB_mPSovermV} we list the values of the mass ratio of
pseudoscalar and vector mesons, $\mps/\mv$, obtained for
$\kval=\ksea$, which is a measure of how heavy our dynamical quark
masses are relative to the real up and down quarks. Given that our
lightest sea quark produces $\mps/\mv\simeq0.67$, we conclude that the
sea quarks used in our simulation are still relatively heavy.

\begin{table}[t]
\begin{center}
\begin{tabular}{|cr@{.}lr@{.}l|}
\hline
\hline
 & \multicolumn{2}{c}{$L=12$}
 & \multicolumn{2}{c|}{$L=16$} \\
$\ksea$ & \multicolumn{2}{c}{$\mps/\mv$}
        & \multicolumn{2}{c|}{$\mps/\mv$} \\
\hline
0.1370 & 0&855\err{2}{3}   & \multicolumn{2}{c|}{\mbox{}} \\
0.1380 & 0&825\err{4}{5}   & \multicolumn{2}{c|}{\mbox{}} \\
0.1390 & 0&785\err{4}{7}   & 0&785\err{6}{7}   \\
0.1395 & 0&710\err{10}{10} & 0&719\err{7}{10}   \\
0.1398 & 0&674\err{9}{23}  & 0&670\err{10}{13} \\
\hline
\hline
\end{tabular}
\end{center}
\caption{The ratio $\mps/\mv$ for $\ksea=\kval$ on the two largest
lattice sizes.
\label{TAB_mPSovermV}}
\end{table}

A useful quantity is the critical value of the hopping parameter,
$\kcrit$. Here, we have used our data for pseudoscalar mesons,
computed for $\kval=\ksea$ and determined $\kcrit$ from a fit
\be
  (a\mps)^2 = aB\,(a\widetilde{m}_q),
\label{EQ_chiralext}
\ee
where
\be
  \widetilde{m}_q=m_q(1+b_{\rm m}\,am_q),\quad
  am_q = \frac{1}{2}\left(\frac{1}{\ksea}-\frac{1}{\kcrit}\right).
\label{EQ_mqtilde}
\ee
Since $b_{\rm m}$ has not been determined non-perturbatively, we have
used its perturbative expression at one loop \cite{SintWeisz97}
\be
   b_{\rm m} = -\frac{1}{2}-0.0962\,g_0^2.
\ee
We considered this choice sufficient for our purposes, given that the
value for $\csw$ used in this study does not completely remove the
leading cutoff effects. For every lattice size we fitted the three
most chiral points to \eq{EQ_chiralext}, and the results for $\kcrit$
are
\bea
  L/a=\,~8: &\quad& \kcrit=0.14047\er{6}{7}  \\
  L/a=12: &\quad& \kcrit=0.14040\er{2}{6}  \\
  L/a=16: &\quad& \kcrit=0.14043\er{2}{3}.
\eea
As an aside we remark that the results for $\kcrit$ obtained using
$b_{\rm m}=0$ in \eq{EQ_mqtilde} are entirely compatible with these
values within errors.

Traditionally, the way to make contact with the physical values of the
light hadron spectrum in quenched simulations has been to extrapolate 
the masses, obtained at several values of $\kappa$, to the chiral
limit. For example, the lattice spacing has usually been determined
by extrapolation of the vector mass and the physical value of $m_\rho$.  
This approach is safe as long as the particles have zero decay width,
as in the quenched approximation, but is no longer feasible in the dynamical
case in which the $\rho$ is not stable. Hence, it desirable to avoid 
extrapolations to the chiral limit, whenever possible.  
With this viewpoint, it has been proposed in \cite{UKQCD_J,APE_a} 
to extract physical values from the region of the strange quark.

\subsection{Finite-size effects in the spectrum}
\label{SEC_had_finvol}

As already mentioned in the introduction, one of the aims of this
study is to acquire experience of the systematics in dynamical
simulations with improved fermions, even if the $O(a)$ effects are not
entirely removed. One important feature to address is the presence of
finite-size effects in the spectrum, which determines the volume at
which we can reliably carry out the calculation. Finite-size effects
in dynamical simulations of hadronic spectrum with Wilson-like
fermions have not been studied in great detail. The only results are
those of the SESAM and T$\chi$L Collaborations with an unimproved
fermionic action, exploring volumes $16^3\cdot 32$ and $24^3\cdot 40$
\cite{SESAM_dyn97}. It is thus important to study and quantify these
effects using an $O(a)$-improved action.

\begin{figure}[t]
\begin{center}
\ewxy{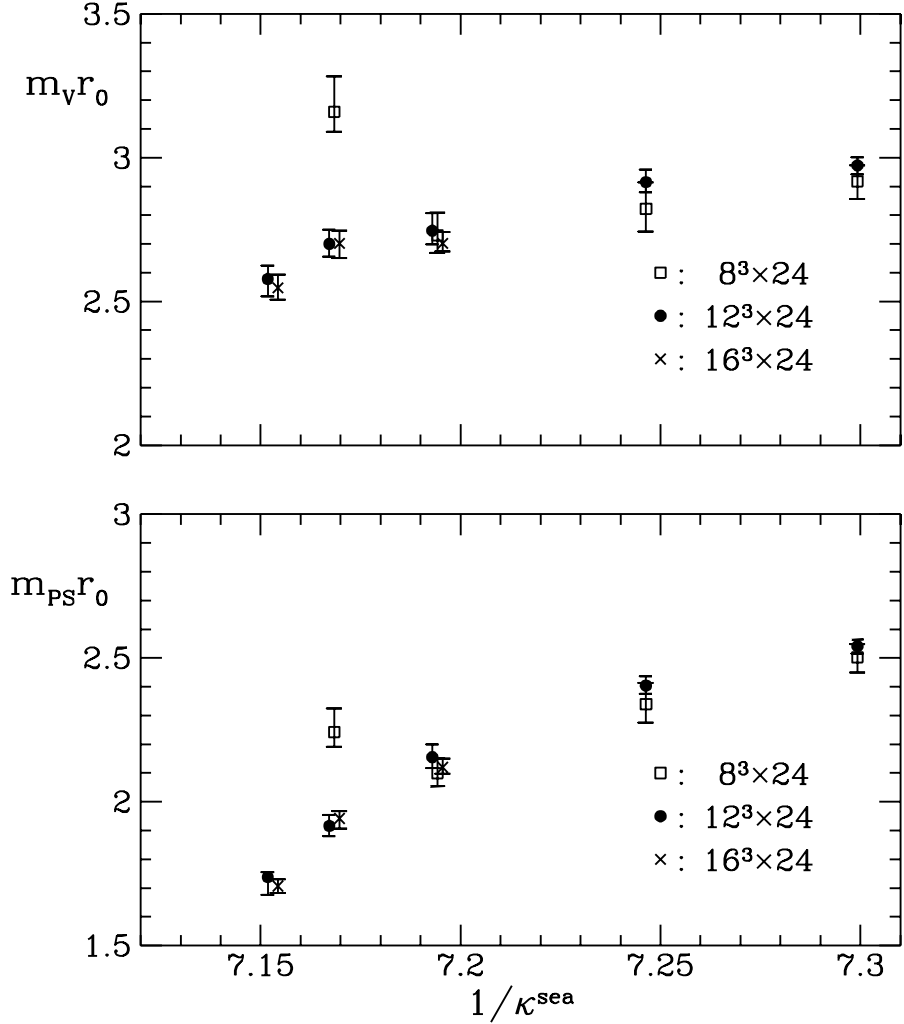}{16cm}
\end{center}
\vspace{-4.5cm}
\caption{Vector and pseudoscalar masses, in units of $r_0$, as a function of 
$1/\ksea=1/\kval$ for different volumes.}
\label{FIG_mesonsr0_vs_k}
\end{figure}

\begin{figure}[t]
\begin{center}
\ewxy{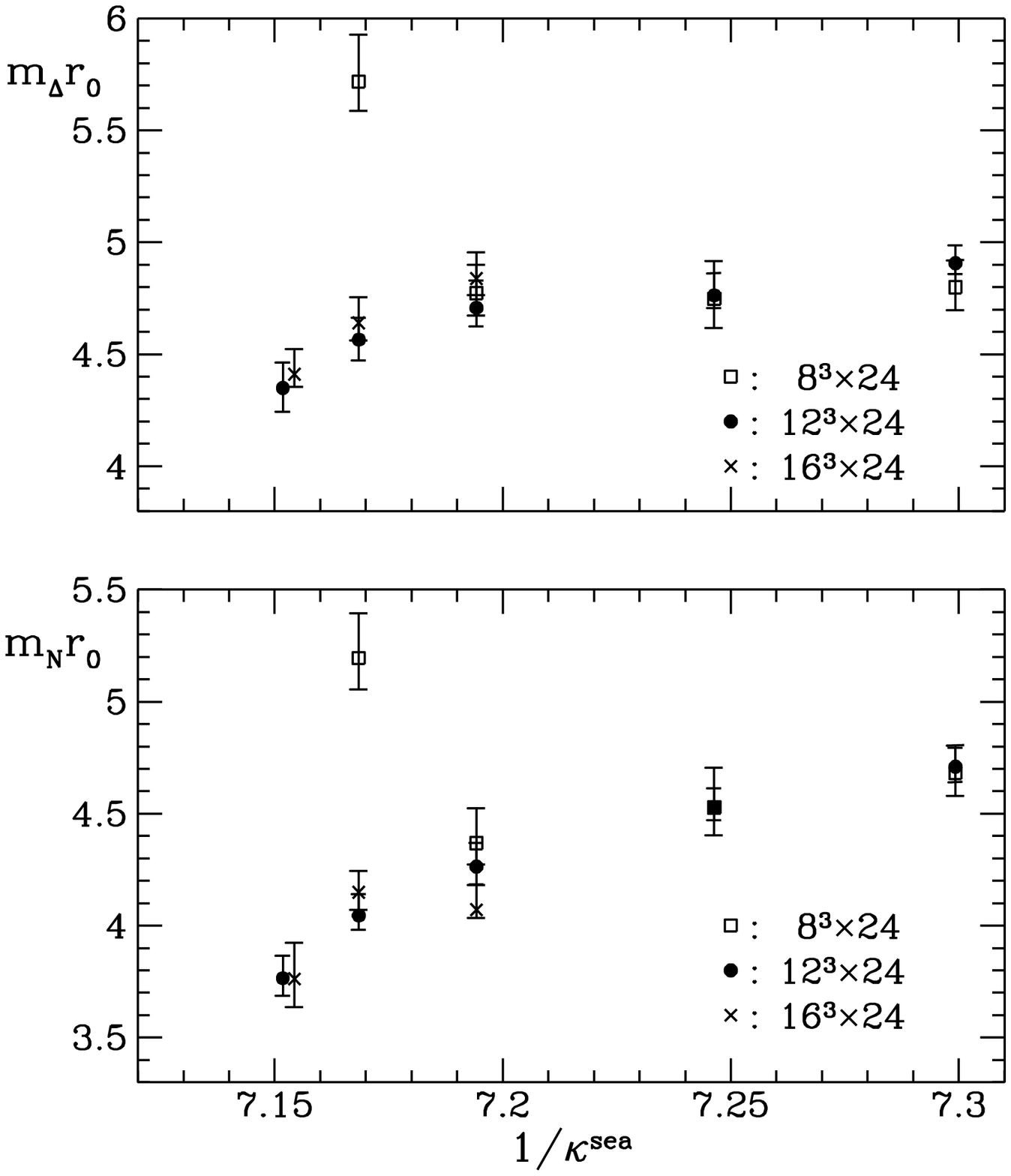}{16cm}
\end{center}
\vspace{-4.5cm}
\caption{$\Delta$ and nucleon masses, in units of $r_0$, as a function of 
$1/\ksea=1/\kval$ for different volumes.}
\label{FIG_baryonsr0_vs_k}
\end{figure}

In Figs.~\ref{FIG_mesonsr0_vs_k} and \ref{FIG_baryonsr0_vs_k} we show
the volume dependence of the meson and baryon masses, in units of
$r_0$, as a function of $1/\ksea$, for $\ksea=\kval$. It confirms the
behaviour already found in the study of the static quark potential.
That is, we find pronounced finite-size effects between $8^3\cdot 24$
and $12^3\cdot 24$, which can grow up to $15-17\%$ in the mesonic
sector and up to $25-28\%$ in the baryonic sector, as we move towards
the most chiral point. On the other hand, between $12^3\cdot 24$ and
$16^3\cdot 24$ we find no significant discrepancy within statistical
accuracy at all common values of the quark mass. From the similarity
of the finite-size behaviour of meson masses and the static quark
potential discussed in section~\ref{SEC_potential}, we conclude that
the bound eq.~(\ref{EQ_bound}) is also valid for simple hadronic
quantities.

\subsection{Sea quark effects in the spectrum}

\begin{figure}[t]
\begin{center}
\vspace{-1.5cm}
\ewxy{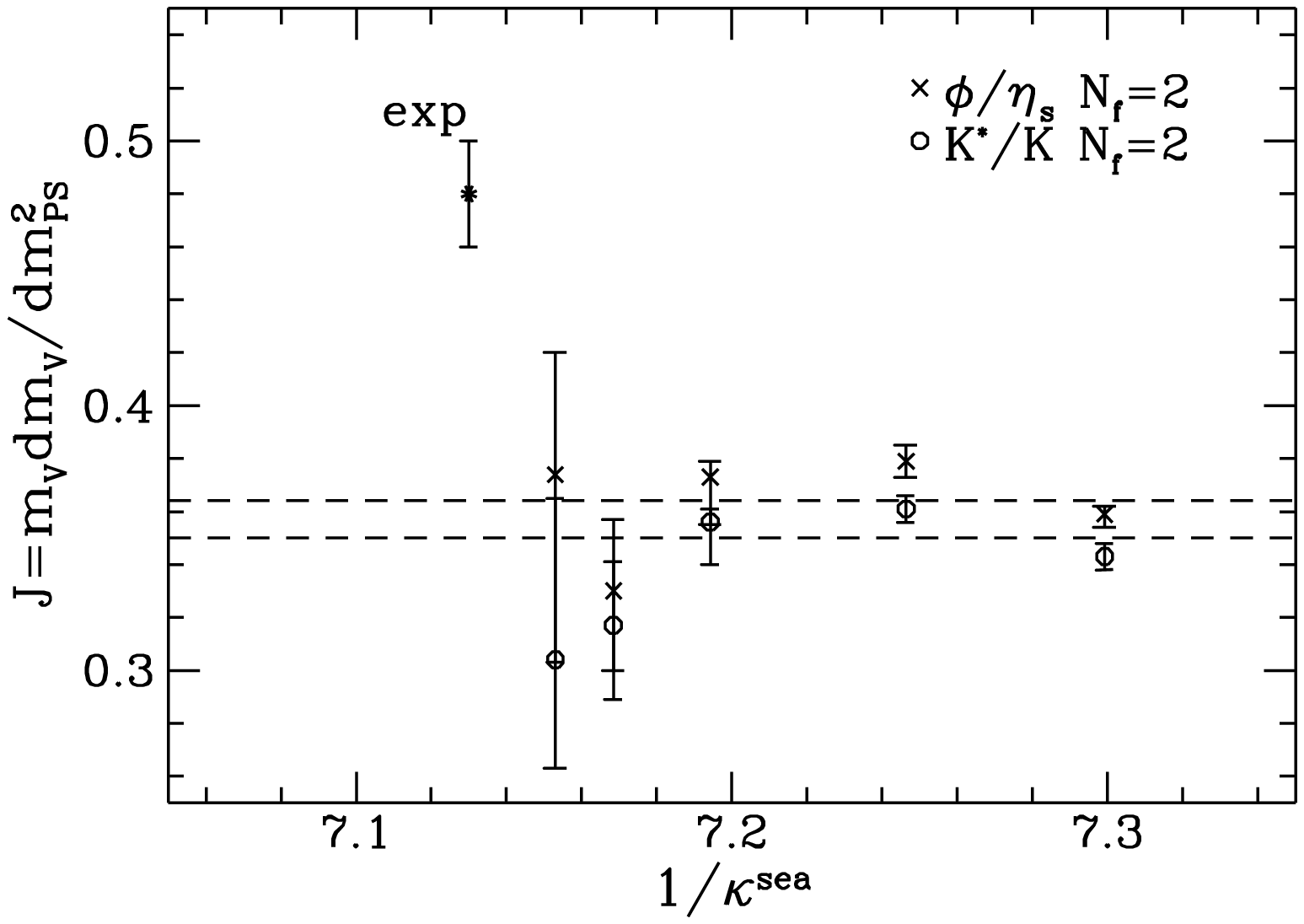}{16cm}
\end{center}
\vspace{-7.0cm}
\caption{Values of the parameter $J$, determined from fixing the
$K^*/K$ and $\phi/\eta_s$, plotted against $1/\ksea$ for $V=12^3\cdot 24$.   
The quenched result, taken from \cite{UKQCD_quenched_TP}, is
represented by the dashed lines.}
\label{FIG_J_vs_k}
\end{figure}

The parameter $J$ has been introduced in \cite{UKQCD_J} as a way to
quantify the discrepancy between the quenched spectrum and experiment.
It is defined as
\begin{equation}
J=m_{K^*}\frac{d{\mv}}{d\mps^2},
\end{equation}
and has the attractive feature that it is dimensionless, does not
involve extrapolations to the chiral limit and is independent of the
quark mass values chosen to evaluate it, provided that $\mv$ depends
linearly on $\mps^2$. The only physical prediction sacrificed to
calculate $J$ is the ratio $m_{K^*}/m_K=1.81$. In the $(\mps^2,{\mv})$
plane this corresponds to a parabola, the intercept of which with the
linear interpolation of the simulated data yields the point
$(m_K^2,m_{K^*})$ of strange mesons. An alternative way to
determine~$J$ is to use mesons with full strange valence quark
content, by assuming that the $\eta_s$ is purely $s\overline{s}$ and
that $m_\phi/m_{\eta_s}=1.57$.

We emphasise that a realistic evaluation of~$J$ in the dynamical case
is not straightforward. As pointed out in \cite{UKQCD_J}, it would not
be appropriate to compare meson masses obtained using different
dynamical quark masses as the lattice spacing $a$ depends on the sea
quark mass. Our approach has been to fix the sea quark mass and for
each sea quark consider different valence quark masses.  Ideally,
since we are looking to interpret the valence quarks as having strange
flavour in the sea of light quarks, we would need to consider values
of $\kval<\ksea$. With our present dataset, this is only possible at
the most chiral sea quark masses. However, since even our lightest sea
quark mass is in the region of that of the strange quark, we do not
expect the values of $J$ to be significantly closer to the
experimental value of $J=0.48(2)$, compared to the quenched
approximation.

In Fig.~\ref{FIG_J_vs_k} we show the values of $J$ and compare them
with the quenched result at $\beta=5.7$ and $V=12^3\cdot 24$
\cite{UKQCD_quenched_TP}. The values of $J$, obtained by either
fixing $m_{K^*}/m_{K}$ or $m_{\phi}/m_{\eta_s}$ show no appreciable
trend towards the experimental point as the sea quark mass decreases.
Moreover, the errors are amplified by the fitting process to extract
the slope ${d{\mv}}/{d\mps^2}$, especially at the lightest $\ksea$.

\begin{figure}[t]
\begin{center}
\vspace{-2.0cm}
\ewxy{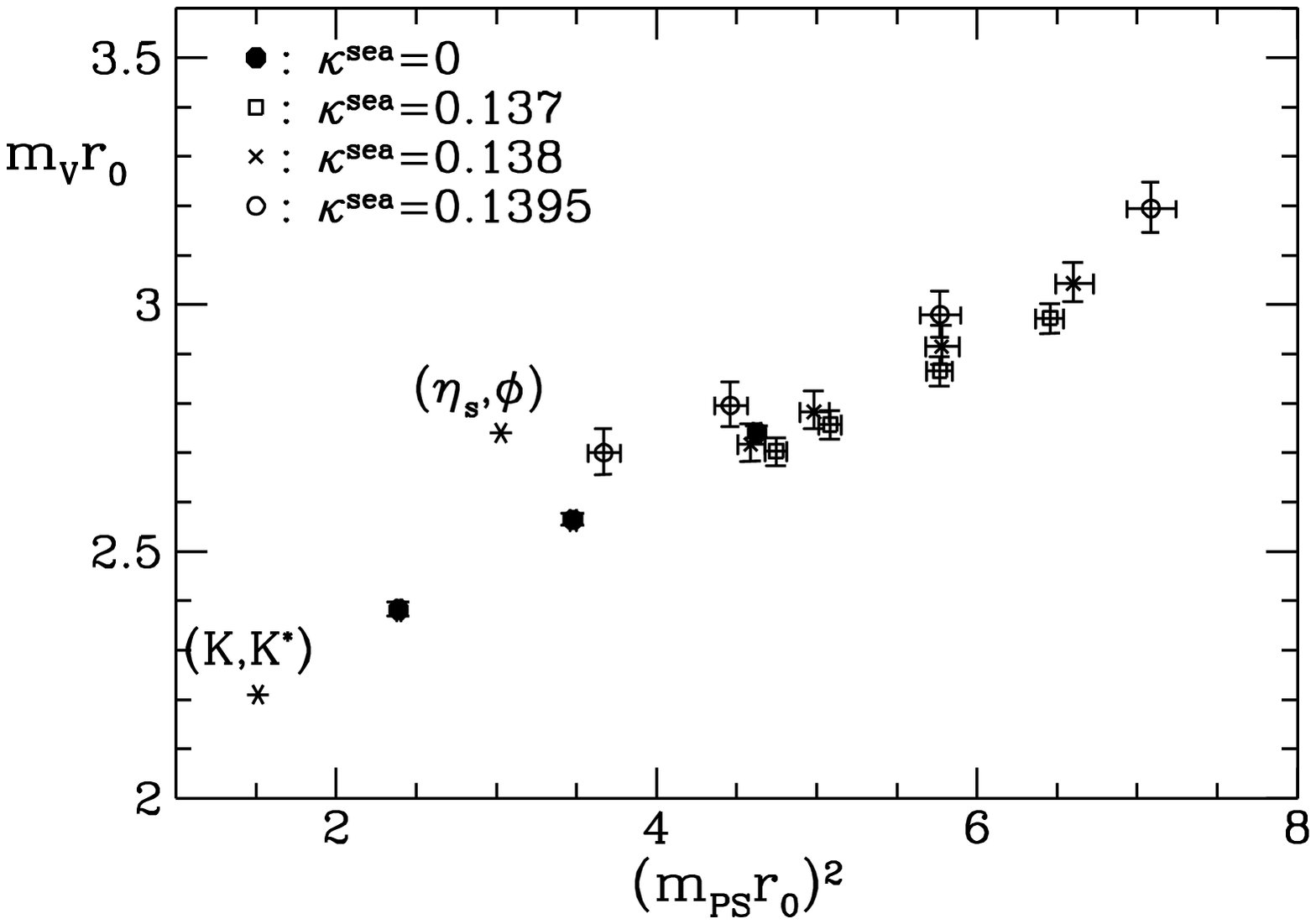}{16cm}
\end{center}
\vspace{-7.0cm}
\caption{Vector mass plotted against the pseudoscalar mass for
$V=12^3\cdot 24$ for several sets of different sea quark masses.
Asterisks denote the experimental points. The quenched results
(labelled $\ksea=0$) are taken from \cite{UKQCD_quenched_TP}.}
\label{FIG_mvr0_vs_mpsr0sq}
\end{figure}

Another way to look for dynamical quark effects in the light hadron
spectrum, which does not involve any fitting, is to concentrate
directly on the plot of the vector mass versus the pseudoscalar mass
squared. This is shown in \fig{FIG_mvr0_vs_mpsr0sq}. As the sea quark
mass is decreased relative to the valence quark mass, one observes a
significant, albeit small trend of the meson masses towards the point
$(m_{\eta_s}^2,m_\phi)$, i.e. the pair of mesons, whose valence quark
content resembles most closely that used in our
simulation. Concentrating on the lighter sea-quark mass,
e.g. $\ksea=0.1395$, we can assert a significant shift compared to the
quenched result, at $\beta=5.7$ and
$V=12^3\cdot24$~\cite{UKQCD_quenched_TP}.

At this stage it is hard to quantify the observed shift, and to
disentangle the genuine sea quark effect from residual lattice
artefacts, which could be fairly large in these simulations. A
suitable approach would be to monitor dynamical quark effects at fixed
lattice spacing. Starting from the quenched approximation, and going
to ever lighter sea quark masses, one would have to perform a sequence
of simulations, which are matched such that they all reproduce the
same value of a suitable lattice scale, e.g. $r_0$ \cite{ISC}. Results
will be published in a future publication, using the fully $O(a)$
improved action for dynamical and quenched
simulations~\cite{UKQCD_dynamical_NP}.

\begin{figure}[h]
\begin{center}
\vspace{-1.5cm}
\ewxy{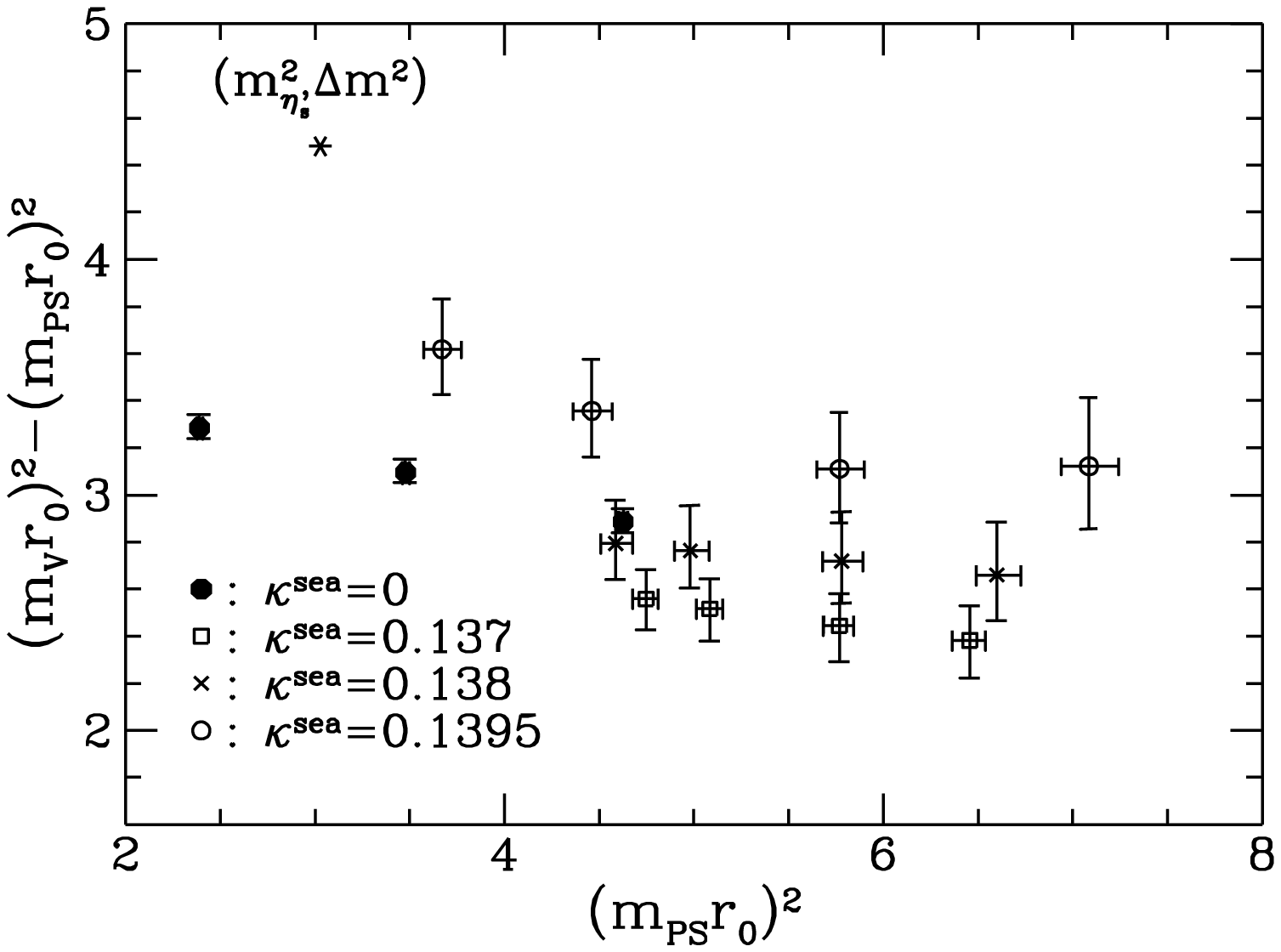}{16cm}
\end{center}
\vspace{-7.0cm}
\caption{Hyperfine splitting $({\mv}r_0)^2-(\mps r_0)^2$ plotted against
$(\mps r_0)^2$ for $V=12^3\cdot 24$.  The quenched
results ($\ksea=0$) are taken from \cite{UKQCD_quenched_TP}.}
\label{FIG_mesonsplit}
\end{figure}

Another quantity, which can be used to highlight the effects of
dynamical quarks, is the vector-pseudoscalar mass splitting. It is
well known that lattice simulations fail to reproduce the experimental
fact that this hyperfine splitting is constant over a large range of
quark masses, $m_{\rm V}^2 - m_{\rm PS}^2\simeq0.55\,\gev^2$. The
discrepancy between the experimental and (much lower) lattice
determinations of $m_{\rm V}^2 - m_{\rm PS}^2$ are partly due to
lattice artefacts \cite{splittings}. However, it is widely expected
that the remaining difference, which becomes more pronounced as the
valence quark mass is increased, is due to dynamical quark effects.
In \fig{FIG_mesonsplit} we plot the meson hyperfine splitting
$({\mv}r_0)^2-(\mps r_0)^2$ versus $(\mps r_0)^2$ for all values of
$\ksea$. The numerical data are compared to the $\phi-\eta_s$
splitting. Despite the relatively poor statistical accuracy it is
obvious that the numerically determined hyperfine splitting shows a
trend towards the experimental point as the sea quark mass is
decreased. As regards the quantification of the relative deviation
from experiment between the results obtained in the quenched
approximation and for finite sea quark mass, the same caveats apply as
for the interpretation of \fig{FIG_mvr0_vs_mpsr0sq}, namely, that the
comparison should be performed for fixed lattice spacing.

\section{Conclusions}
\label{SEC_concl}

In this paper, we have presented the first results on the light hadron
spectrum and the static quark potential obtained from dynamical
simulations using an $O(a)$ improved fermion action at
$\beta=5.2$. Sea quark masses were chosen such that $\mps/\mv$ was in
the range~$0.86$ down to~$0.67$. The value of $\csw=1.76$ was not
appropriate to remove all leading discretisation effects. We wish to
point out, however, that at such a low $\beta$ value residual lattice
artefacts could be relatively large, even after full $O(a)$
improvement. This question clearly requires further investigation.

We have addressed the important issue of finite-size effects, which
are expected to be larger in dynamical simulations, compared to the
quenched approximation. By simulating three different lattice sizes
for a range of sea quark masses, we found that 
finite-size effects are practically absent for box sizes
$L\,\,\gtaeq\,\,1.6\,\fm$, and sea quark masses which give
$\mps/\mv\,\,\gtaeq\,\,0.67$. This is observed in the data for the
static quark potential and hadron masses. However, for lighter
dynamical quarks one would expect that yet larger volumes are
required. 

Instead of presenting quantitative results for the light hadron
spectrum, we have concentrated on highlighting the effects of
dynamical quarks. Although the evidence for string breaking in the
data for the static quark potential and for an improved behaviour of
the parameter~$J$ remains inconclusive, we have detected significant
effects due to dynamical quarks. Namely, the Coulombic part of the
static quark potential is enhanced for finite sea quark mass, and the
vector-pseudoscalar hyperfine splitting moves closer to the
experimental value as the sea quark mass is decreased. Furthermore,
pairs of $(\mv r_0,(\mps r_0)^2)$ show a trend towards experiment when
dynamical quarks are ``switched on''.

The results presented here serve as a guideline for ongoing
investigations, which are performed using the fully $O(a)$ improved
action for two flavours of dynamical quarks, whose masses are closer
to the chiral limit than those used in this study.

\section*{Acknowledgements}
 
We are grateful to the Alpha Collaboration for communicating to us
their preliminary value for the improvement coefficient at
$\beta=5.2$.
We acknowledge the support of the Particle Physics \& Astronomy
Research Council under grant GR/L22744.
H.W. acknowledges the support of PPARC through the award of an
Advanced Fellowship.

\newpage
\appendix
\section{Tables}
\label{APP_tables}

Here we list the meson and baryon masses for all combinations of
$\ksea$ and $\kval$, and on all lattice sizes used in the simulations.

In the following tables we also list the fitting interval and the
value of $\chi^2/\rm d.o.f.$, obtained from a correlated fit to
eqs.~(\ref{EQ_fitmeson}) or~(\ref{EQ_fitbaryon}), respectively. Some
of the fits produce values of $\chi^2/\rm d.o.f.$, which would
normally be regarded as unacceptably large, even though we are
confident that the fit intervals were properly tuned. In such cases we
have compared the results for the masses to those obtained using an
uncorrelated fit over the same interval. We always found that the
uncorrelated fit gave results which, within errors, were perfectly
compatible with those from the correlated fit, whilst producing values
for $\chi^2/\rm d.o.f.$, which were significantly below one. Thus we
are confident that the results presented in the tables below are
reliable, that a double exponential is an appropriate model function
in all cases, but that correlations in the data are clearly present.

\setlength{\textheight}{245mm} 
\begin{table}[h]
\begin{center}
\begin{tabular}{|ccccccc|}\hline\hline
$L^3\cdot T$ & $\kappa^{sea}$ & $\kappa^{val}$ & ${\mps}a$ & $\mps r_0$ & Fit & $\chi^2 /$ dof  \\ \hline \hline
$8^3\cdot 24$ & 0.1370 & 0.1370 &  1.119\err{  8}{  4} &  2.502\err{ 47}{ 52} & [ 6,11] & 12.70 / 6  \\
              &        & 0.1380 &  1.059\err{  8}{  4} &  2.368\err{ 45}{ 50} & [ 6,11] & 12.77 / 6  \\
              &        & 0.1390 &  0.997\err{  9}{  5} &  2.229\err{ 44}{ 47} & [ 6,11] & 13.20 / 6  \\
              &        & 0.1395 &  0.965\err{  9}{  4} &  2.158\err{ 43}{ 45} & [ 6,11] & 13.57 / 6  \\ \hline 
$8^3\cdot 24$ & 0.1380 & 0.1370 &  1.011\err{ 10}{  2} &  2.502\err{ 79}{ 68} & [ 5,11] & 14.94 / 8  \\
              &        & 0.1380 &  0.945\err{ 11}{  2} &  2.339\err{ 75}{ 64} & [ 5,11] & 13.28 / 8  \\
              &        & 0.1390 &  0.874\err{ 12}{  6} &  2.164\err{ 72}{ 61} & [ 5,11] &  7.26 / 8  \\
              &        & 0.1395 &  0.840\err{ 14}{  6} &  2.078\err{ 71}{ 58} & [ 5,11] &  7.50 / 8  \\ \hline 
$8^3\cdot 24$ & 0.1390 & 0.1370 &  0.870\err{  9}{  5} &  2.514\err{ 58}{ 53} & [ 6,11] &  8.14 / 6  \\
              &        & 0.1380 &  0.799\err{ 10}{  5} &  2.309\err{ 57}{ 49} & [ 6,11] &  8.18 / 6  \\
              &        & 0.1390 &  0.726\err{ 11}{  6} &  2.099\err{ 53}{ 45} & [ 5,11] & 10.24 / 8  \\
              &        & 0.1395 &  0.686\err{ 12}{  5} &  1.982\err{ 53}{ 43} & [ 5,11] & 10.86 / 8  \\ \hline 
$8^3\cdot 24$ & 0.1395 & 0.1370 &  0.786\err{  9}{ 14} &  2.924\err{ 78}{ 78} & [ 6,11] &  9.70 / 6  \\
              &        & 0.1380 &  0.710\err{  9}{ 12} &  2.640\err{ 72}{ 68} & [ 6,11] & 10.75 / 6  \\
              &        & 0.1390 &  0.642\err{ 15}{ 10} &  2.387\err{ 80}{ 60} & [ 5,11] & 15.07 / 8  \\
              &        & 0.1395 &  0.603\err{ 17}{  7} &  2.242\err{ 82}{ 51} & [ 4,11] & 17.30 / 10 \\ \hline 
\hline
\end{tabular}
\end{center}
\label{TAB_pseudo_V8X24}
\caption{Pseudoscalar masses on $8^3\cdot24$, in lattice units and in
units of $r_0$.}
\end{table}

\begin{table}
\begin{center}
\begin{tabular}{|ccccccc|}\hline\hline
$L^3\cdot T$ & $\kappa^{sea}$ & $\kappa^{val}$ & ${\mps}a$ & $\mps r_0$ & Fit & $\chi^2 /$ dof  \\ \hline \hline
$12^3\cdot 24$ & 0.1370 & 0.1370 &  1.108\err{  3}{  2} &  2.541\err{ 23}{ 26} & [ 6,11] &  9.93 / 6  \\
               &  & 0.1380 &  1.047\err{  3}{  2} &  2.402\err{ 22}{ 25} & [ 5,11] & 11.18 / 8  \\
               &  & 0.1390 &  0.983\err{  4}{  2} &  2.255\err{ 21}{ 23} & [ 5,11] & 11.65 / 8  \\
               &  & 0.1395 &  0.950\err{  4}{  3} &  2.179\err{ 21}{ 23} & [ 5,11] & 11.83 / 8  \\ \hline 
$12^3\cdot 24$ & 0.1380 & 0.1370 &  1.000\err{  3}{  2} &  2.569\err{ 35}{ 30} & [ 6,11] & 10.41 / 6  \\
               &  & 0.1380 &  0.936\err{  3}{  2} &  2.404\err{ 33}{ 29} & [ 6,11] & 10.79 / 6  \\
               &  & 0.1390 &  0.869\err{  4}{  2} &  2.232\err{ 31}{ 27} & [ 6,11] & 11.37 / 6  \\
               &  & 0.1395 &  0.834\err{  4}{  2} &  2.142\err{ 30}{ 26} & [ 6,11] & 11.56 / 6  \\ \hline 
$12^3\cdot 24$ & 0.1390 & 0.1370 &  0.858\err{  4}{  3} &  2.613\err{ 51}{ 45} & [ 7,11] &  5.93 / 4  \\
               &  & 0.1380 &  0.785\err{  4}{  3} &  2.391\err{ 48}{ 42} & [ 7,11] &  4.16 / 4  \\
               &  & 0.1390 &  0.707\err{  5}{  3} &  2.155\err{ 44}{ 38} & [ 7,11] &  2.69 / 4  \\
               &  & 0.1395 &  0.669\err{  5}{  3} &  2.037\err{ 42}{ 36} & [ 6,11] & 16.86 / 6  \\ \hline 
$12^3\cdot 24$ & 0.1395 & 0.1370 &  0.775\err{  6}{  4} &  2.662\err{ 42}{ 39} & [ 7,11] & 21.46 / 4  \\
               &  & 0.1380 &  0.699\err{  5}{  4} &  2.402\err{ 38}{ 36} & [ 6,11] & 26.14 / 6  \\
               &  & 0.1390 &  0.615\err{  6}{  4} &  2.112\err{ 36}{ 32} & [ 6,11] & 19.70 / 6  \\
               &  & 0.1395 &  0.558\err{  8}{  7} &  1.916\err{ 38}{ 35} & [ 5,11] &  8.96 / 8  \\ \hline 
$12^3\cdot 24$ & 0.1398 & 0.1380 &  0.639\err{  5}{  4} &  2.332\err{ 27}{ 21} & [ 7,11] &  3.63 / 4  \\
               &  & 0.1390 &  0.551\err{  4}{  9} &  2.011\err{ 23}{ 37} & [ 7,11] &  2.09 / 4  \\
               &  & 0.1395 &  0.492\err{  3}{  9} &  1.797\err{ 18}{ 36} & [ 6,11] &  3.41 / 6  \\
               &  & 0.1398 &  0.476\err{  3}{ 16} &  1.738\err{ 17}{ 61} & [ 6,11] &  8.14 / 6  \\ \hline 
\hline
\end{tabular}
\end{center}
\label{TAB_pseudo_V12X24}
\caption{Pseudoscalar masses on $12^3\cdot24$, in lattice units and in
units of $r_0$.} 
\end{table}

\begin{table}
\begin{center}
\begin{tabular}{|ccccccc|}\hline\hline
$L^3\cdot T$ & $\kappa^{sea}$ & $\kappa^{val}$ & ${\mps}a$ & $\mps r_0$ & Fit & $\chi^2 /$ dof  \\ \hline \hline
$16^3\cdot 24$ & 0.1390 & 0.1390 &  0.701\er{  6}{  5} &  2.120\err{ 30}{ 22} & [ 6,11] & 10.62 / 6 \\
               &  & 0.1395 &  0.660\er{  6}{  4} &  1.998\err{ 28}{ 21} & [ 6,11] &  9.31 / 6  \\
               &  & 0.1398 &  0.635\er{  6}{  5} &  1.922\err{27}{ 22} & [ 6,11] &  8.34 / 6  \\ \hline
$16^3\cdot 24$ & 0.1395 & 0.1390 &  0.610\er{  4}{  2} &  2.101\err{ 28}{ 36} & [ 6,11] &  7.04 / 6  \\
               &  & 0.1395 &  0.564\er{  3}{  4} &  1.942\err{ 25}{ 36} & [ 6,11] &  3.05 / 6  \\
               &  & 0.1398 &  0.537\er{  4}{  3} &  1.848\err{ 26}{ 32} & [ 7,11] &  2.99 / 4  \\ \hline 
$16^3\cdot 24$ & 0.1398 & 0.1390 &  0.551\er{  5}{  5} &  2.011\err{ 26}{ 26} & [ 7,11] & 10.66 / 4  \\
               &  & 0.1395 &  0.502\er{  5}{  6} &  1.834\err{ 24}{ 28} & [ 7,11] &  8.48 / 4  \\
               &  & 0.1398 &  0.468\er{  5}{  5} &  1.707\err{ 23}{ 24} & [ 7,11] &  6.02 / 4  \\ \hline 
\hline
\end{tabular}
\end{center}
\label{TAB_pseudo_V16X24}
\caption{Pseudoscalar masses on $16^3\cdot24$, in lattice units and in
units of $r_0$.} 
\end{table}

\begin{table}
\begin{center}
\begin{tabular}{|ccccccc|}\hline\hline
$L^3\cdot T$ & $\kappa^{sea}$ & $\kappa^{val}$ & ${\mv}a$ & ${\mv}r_0$ & Fit & $\chi^2 /$ dof  \\ \hline \hline
$8^3\cdot 24$ & 0.1370 & 0.1370 &  1.305\err{ 11}{  6} &  2.917\errr{ 57}{ 61} & [ 7,11] &  5.82 / 4  \\
              &  & 0.1380 &  1.257\err{ 12}{  6} &  2.810\errr{ 56}{ 59} & [ 7,11] &  5.09 / 4  \\
              &  & 0.1390 &  1.209\err{ 14}{  7} &  2.703\errr{ 56}{ 58} & [ 7,11] &  5.27 / 4  \\
              &  & 0.1395 &  1.185\err{ 14}{  8} &  2.650\errr{ 56}{ 57} & [ 7,11] &  5.78 / 4  \\ \hline 
$8^3\cdot 24$ & 0.1380 & 0.1370 &  1.190\err{ 14}{  7} &  2.946\errr{ 95}{ 82} & [ 6,11] &  7.30 / 6  \\
              &  & 0.1380 &  1.140\err{ 15}{  8} &  2.822\errr{ 92}{ 79} & [ 6,11] &  7.68 / 6  \\
              &  & 0.1390 &  1.090\err{ 16}{  9} &  2.697\errr{ 90}{ 76} & [ 6,11] &  9.02 / 6  \\
              &  & 0.1395 &  1.064\err{ 17}{ 10} &  2.634\errr{ 89}{ 76} & [ 6,11] & 10.05 / 6  \\ \hline 
$8^3\cdot 24$ & 0.1390 & 0.1370 &  1.045\err{ 13}{  7} &  3.022\errr{ 72}{ 64} & [ 7,11] &  5.21 / 4  \\
              &  & 0.1380 &  0.998\err{ 16}{  9} &  2.885\errr{ 76}{ 64} & [ 6,11] &  6.49 / 6  \\
              &  & 0.1390 &  0.944\err{ 19}{ 10} &  2.730\errr{ 79}{ 61} & [ 6,11] &  4.66 / 6  \\
              &  & 0.1395 &  0.918\err{ 20}{ 10} &  2.653\errr{ 80}{ 61} & [ 6,11] &  4.08 / 6  \\ \hline 
$8^3\cdot 24$ & 0.1395 & 0.1370 &  0.961\err{ 30}{  9} &  3.573\errr{141}{ 77} & [ 7,11] & 12.79 / 4  \\
              &  & 0.1380 &  0.917\err{ 24}{  6} &  3.411\errr{122}{ 70} & [ 6,11] & 11.80 / 6  \\
              &  & 0.1390 &  0.874\err{ 34}{  5} &  3.251\errr{149}{ 65} & [ 6,11] & 11.01 / 6  \\
              &  & 0.1395 &  0.850\err{ 26}{  9} &  3.159\errr{124}{ 69} & [ 6,11] & 11.17 / 6  \\ \hline 
\hline
\end{tabular}
\end{center}
\label{TAB_vector_V8X24}
\caption{Vector masses on $8^3\cdot24$, in lattice units and in units
of $r_0$.} 
\end{table}

\begin{table}
\begin{center}
\begin{tabular}{|ccccccc|}\hline\hline
$L^3\cdot T$ & $\kappa^{sea}$ & $\kappa^{val}$ & ${\mv}a$ & ${\mv}r_0$ & Fit & $\chi^2 /$ dof  \\ \hline \hline
$12^3\cdot 24$ & 0.1370 & 0.1370 &  1.296\err{  6}{  4} &  2.973\err{ 29}{ 31} & [ 7,11] &  3.85 / 4  \\
               &  & 0.1380 &  1.249\err{  6}{  5} &  2.866\err{ 28}{ 31} & [ 7,11] &  4.17 / 4  \\
               &  & 0.1390 &  1.202\err{  6}{  5} &  2.757\err{ 28}{ 30} & [ 7,11] &  4.77 / 4  \\
               &  & 0.1395 &  1.178\err{  7}{  5} &  2.703\err{ 28}{ 29} & [ 7,11] &  5.19 / 4  \\ \hline 
$12^3\cdot 24$ & 0.1380 & 0.1370 &  1.185\err{  6}{  4} &  3.043\err{ 43}{ 37} & [ 7,11] & 24.88 / 4  \\
               &  & 0.1380 &  1.135\err{  7}{  4} &  2.915\err{ 43}{ 36} & [ 6,11] & 33.99 / 6  \\
               &  & 0.1390 &  1.084\err{  8}{  4} &  2.783\err{ 42}{ 34} & [ 6,11] & 28.64 / 6  \\
               &  & 0.1395 &  1.058\err{  9}{  4} &  2.717\err{ 42}{ 34} & [ 6,11] & 25.81 / 6  \\ \hline 
$12^3\cdot 24$ & 0.1390 & 0.1370 &  1.016\err{  4}{  5} &  3.094\err{ 60}{ 55} & [ 6,11] &  6.99 / 6  \\
               &  & 0.1380 &  0.962\err{  7}{  3} &  2.931\err{ 59}{ 51} & [ 7,11] & 14.24 / 4  \\
               &  & 0.1390 &  0.901\err{ 10}{  3} &  2.746\err{ 61}{ 48} & [ 7,11] &  8.29 / 4  \\
               &  & 0.1395 &  0.891\err{  9}{  1} &  2.714\err{ 59}{ 47} & [ 6,11] & 36.73 / 6  \\ \hline 
$12^3\cdot 24$ & 0.1395 & 0.1370 &  0.930\err{  9}{  7} &  3.195\err{ 54}{ 49} & [ 7,11] & 29.92 / 4  \\
               &  & 0.1380 &  0.867\err{  7}{  6} &  2.980\err{ 48}{ 46} & [ 6,11] & 12.76 / 6  \\
               &  & 0.1390 &  0.814\err{  8}{  6} &  2.796\err{ 48}{ 43} & [ 5,11] & 13.67 / 8  \\
               &  & 0.1395 &  0.786\err{  9}{  7} &  2.700\err{ 49}{ 44} & [ 5,11] & 13.10 / 8  \\ \hline 
$12^3\cdot 24$ & 0.1398 & 0.1380 &  0.791\err{ 16}{  1} &  2.890\err{ 64}{ 20} & [ 7,11] & 25.71 / 4  \\
               &  & 0.1390 &  0.735\err{  7}{ 14} &  2.685\err{ 33}{ 54} & [ 6,11] & 14.22 / 6  \\
               &  & 0.1395 &  0.725\err{ 12}{ 16} &  2.647\err{ 49}{ 62} & [ 5,11] & 29.11 / 8  \\
               &  & 0.1398 &  0.706\err{ 12}{ 16} &  2.578\err{ 47}{ 61} & [ 5,11] & 25.26 / 8  \\ \hline 
\hline
\end{tabular}
\end{center}
\label{TAB_vector_V12X24}
\caption{Vector masses on $12^3\cdot24$, in lattice units and in units
of $r_0$.} 
\end{table}

\begin{table}
\begin{center}
\begin{tabular}{|ccccccc|}\hline\hline
$L^3\cdot T$ & $\kappa^{sea}$ & $\kappa^{val}$ & ${\mv}a$ & ${\mv}r_0$ & Fit & $\chi^2 /$ dof  \\ \hline \hline
$16^3\cdot 24$ & 0.1390 & 0.1390 &  0.893\err{  9}{  6} &  2.702\err{ 39}{ 28} & [ 7,11] & 14.64 / 4  \\
               &  & 0.1395 &  0.863\err{ 10}{  6} &  2.610\err{ 40}{ 28} & [ 7,11] & 14.79 / 4  \\
               &  & 0.1398 &  0.852\err{ 11}{  6} &  2.580\err{ 43}{ 27} & [ 4,11] & 22.82 / 10 \\ \hline 
$16^3\cdot 24$ & 0.1395 & 0.1390 &  0.813\err{  8}{  6} &  2.801\err{ 43}{ 51} & [ 7,11] & 13.43 / 4  \\
               &  & 0.1395 &  0.785\err{  9}{  7} &  2.702\err{ 45}{ 51} & [ 7,11] & 12.70 / 4  \\
               &  & 0.1398 &  0.773\err{ 10}{  8} &  2.664\err{ 46}{ 53} & [ 5,11] & 14.00 / 8  \\ \hline 
$16^3\cdot 24$ & 0.1398 & 0.1390 &  0.749\err{  9}{  8} &  2.733\err{ 41}{ 37} & [ 7,11] &  2.09 / 4  \\
               &  & 0.1395 &  0.716\err{ 10}{  8} &  2.614\err{ 44}{ 37} & [ 7,11] &  1.30 / 4 \\ 
               &  & 0.1398 &  0.698\err{ 11}{ 10} &  2.547\err{ 47}{ 41} & [ 7,11] &  1.60 / 4  \\ \hline 
\hline
\end{tabular}
\end{center}
\label{TAB_vector_V16X24}
\caption{Vector masses on $16^3\cdot24$, in lattice units and in units
of $r_0$.} 
\end{table}

\begin{table}
\begin{center}
\begin{tabular}{|ccccccc|}\hline\hline
$L^3\cdot T$ & $\kappa^{sea}$ & $\kappa^{val}$ & $m_{N}a$ & $m_{N}r_0$ & Fit & $\chi^2 /$ dof  \\ \hline \hline
$8^3\cdot 24$ & 0.1370 & 0.1370 &  2.094\err{42}{ 14} &  4.681\errr{124}{102} & [ 6,11] & 14.59 / 6  \\
              &  & 0.1380 &  2.020\err{43}{ 15} &  4.517\errr{124}{ 99} & [ 6,11] & 13.29 / 6  \\
              &  & 0.1390 &  1.945\err{44}{ 18} &  4.349\errr{124}{ 98} & [ 6,11] & 12.07 / 6  \\
              &  & 0.1395 &  1.907\err{49}{ 21} &  4.264\errr{132}{ 99} & [ 6,11] & 11.54 / 6  \\ \hline 
$8^3\cdot 24$ & 0.1380 & 0.1370 &  1.926\err{40}{ 14} &  4.768\errr{174}{134} & [ 6,11] & 11.13 / 6  \\
              &  & 0.1380 &  1.830\err{47}{ 10} &  4.528\errr{178}{125} & [ 6,11] &  9.43 / 6  \\
              &  & 0.1390 &  1.730\err{55}{  8} &  4.282\errr{188}{117} & [ 6,11] &  7.65 / 6  \\
              &  & 0.1395 &  1.677\err{63}{  7} &  4.151\errr{198}{114} & [ 6,11] &  6.74 / 6  \\ \hline 
$8^3\cdot 24$ & 0.1390 & 0.1370 &  1.687\err{29}{ 12} &  4.877\errr{132}{103} & [ 6,11] & 11.89 / 6  \\
              &  & 0.1380 &  1.596\err{38}{ 10} &  4.613\errr{146}{ 97} & [ 6,11] &  9.77 / 6  \\
              &  & 0.1390 &  1.511\err{44}{ 14} &  4.368\errr{155}{ 96} & [ 6,11] &  8.65 / 6  \\
              &  & 0.1395 &  1.473\err{47}{ 20} &  4.259\errr{162}{103} & [ 6,11] &  8.89 / 6  \\ \hline 
$8^3\cdot 24$ & 0.1395 & 0.1370 &  1.606\err{25}{ 22} &  5.971\errr{171}{141} & [ 6,11] &  8.73 / 6  \\
              &  & 0.1380 &  1.535\err{24}{ 26} &  5.706\errr{164}{148} & [ 6,11] &  7.84 / 6  \\
              &  & 0.1390 &  1.463\err{34}{ 48} &  5.440\errr{181}{209} & [ 7,11] &  4.81 / 4  \\
              &  & 0.1395 &  1.397\err{42}{ 26} &  5.195\errr{200}{140} & [ 4,11] &  9.34 / 100 \\ \hline 
\hline
\end{tabular}
\end{center}
\label{TAB_nucleon_V8X24}
\caption{Nucleon masses on $8^3\cdot24$, in lattice units and in units
of $r_0$.} 
\end{table}

\begin{table}
\begin{center}
\begin{tabular}{|ccccccc|}\hline\hline
$L^3\cdot T$ & $\kappa^{sea}$ & $\kappa^{val}$ & $m_{N}a$ & $m_{N}r_0$ & Fit & $\chi^2 /$ dof  \\ \hline \hline
$12^3\cdot 24$ & 0.1370 & 0.1370 &  2.053\err{ 32}{ 22} &  4.710\errr{ 85}{ 69} & [ 7,11] &  8.29 / 4  \\
               &  & 0.1380 &  1.975\err{ 36}{ 19} &  4.531\errr{ 91}{ 62} & [ 7,11] &  7.97 / 4  \\
               &  & 0.1390 &  1.900\err{ 33}{ 10} &  4.359\errr{ 85}{ 50} & [ 6,11] & 10.42 / 6  \\
               &  & 0.1395 &  1.860\err{ 33}{ 11} &  4.267\errr{ 84}{ 49} & [ 6,11] &  9.47 / 6  \\ \hline 
$12^3\cdot 24$ & 0.1380 & 0.1370 &  1.850\err{ 24}{  9} &  4.752\errr{ 88}{ 60} & [ 7,11] &  4.95 / 4  \\
               &  & 0.1380 &  1.763\err{ 23}{ 10} &  4.528\errr{ 85}{ 58} & [ 7,11] &  4.85 / 4  \\
               &  & 0.1390 &  1.673\err{ 24}{  9} &  4.296\errr{ 83}{ 56} & [ 7,11] &  5.56 / 4  \\
               &  & 0.1395 &  1.626\err{ 24}{ 11} &  4.176\errr{ 83}{ 56} & [ 7,11] &  6.19 / 4  \\ \hline 
$12^3\cdot 24$ & 0.1390 & 0.1370 &  1.604\err{ 20}{  9} &  4.886\errr{111}{ 88} & [ 7,11] & 15.34 / 4  \\
               &  & 0.1380 &  1.505\err{ 21}{ 10} &  4.585\errr{107}{ 83} & [ 7,11] & 13.01 / 4  \\
               &  & 0.1390 &  1.399\err{ 22}{ 11} &  4.263\errr{106}{ 80} & [ 7,11] & 10.61 / 4  \\
               &  & 0.1395 &  1.343\err{ 24}{ 11} &  4.089\errr{107}{ 77} & [ 7,11] &  9.33 / 4  \\ \hline 
$12^3\cdot 24$ & 0.1395 & 0.1370 &  1.434\err{ 19}{ 10} &  4.927\errr{ 95}{ 76} & [ 7,11] & 13.69 / 4  \\
               &  & 0.1380 &  1.335\err{ 20}{ 11} &  4.585\errr{ 94}{ 73} & [ 7,11] & 10.29 / 4  \\
               &  & 0.1390 &  1.233\err{ 20}{ 12} &  4.234\errr{ 92}{ 71} & [ 7,11] &  7.69 / 4  \\
               &  & 0.1395 &  1.178\err{ 22}{ 10} &  4.045\errr{ 95}{ 65} & [ 6,11] &  5.05 / 6  \\ \hline 
$12^3\cdot 24$ & 0.1398 & 0.1380 &  1.236\err{ 26}{ 15} &  4.514\errr{102}{ 63} & [ 6,11] & 15.02 / 6  \\
               &  & 0.1390 &  1.124\err{ 23}{ 23} &  4.106\errr{ 90}{ 89} & [ 7,11] & 14.30 / 4  \\
               &  & 0.1395 &  1.069\err{ 25}{ 20} &  3.902\errr{ 97}{ 76} & [ 6,11] & 14.69 / 6  \\
               &  & 0.1398 &  1.031\err{ 26}{ 20} &  3.765\errr{101}{ 78} & [ 6,11] & 14.74 / 6  \\ \hline 
\hline
\end{tabular}
\end{center}
\label{TAB_nucleon_V12X24}
\caption{Nucleon masses on $12^3\cdot24$, in lattice units and in
units of $r_0$.} 
\end{table}

\begin{table}
\begin{center}
\begin{tabular}{|ccccccc|}\hline\hline
$L^3\cdot T$ & $\kappa^{sea}$ & $\kappa^{val}$ & $m_{N}a$ & $m_{N}r_0$ & Fit & $\chi^2 /$ dof  \\ \hline \hline
$16^3\cdot 24$ & 0.1390 & 0.1390 &  1.345\err{ 33}{  5} &  4.071\errr{110}{ 36} & [ 7,11] &  5.49 / 4  \\
               &  & 0.1395 &  1.295\err{ 32}{  6} &  3.918\errr{106}{ 36} & [ 7,11] &  6.24 / 4  \\
               &  & 0.1398 &  1.264\err{ 31}{  7} &  3.826\errr{102}{ 37} & [ 7,11] &  6.72 / 4  \\ \hline 
$16^3\cdot 24$ & 0.1395 & 0.1390 &  1.261\err{ 25}{ 10} &  4.344\errr{ 99}{ 80} & [ 7,11] & 13.78 / 4  \\
               &  & 0.1395 &  1.205\err{ 24}{ 11} &  4.149\errr{ 96}{ 79} & [ 7,11] & 10.92 / 4  \\
               &  & 0.1398 &  1.164\err{ 24}{ 19} &  4.010\errr{ 96}{ 92} & [ 7,11] &  9.08 / 4  \\ \hline 
$16^3\cdot 24$ & 0.1398 & 0.1390 &  1.144\err{ 36}{ 31} &  4.178\errr{136}{117} & [ 7,11] & 23.51 / 4  \\
               &  & 0.1395 &  1.065\err{ 47}{ 35} &  3.888\errr{173}{133} & [ 7,11] & 19.78 / 4  \\
               &  & 0.1398 &  1.030\err{ 43}{ 34} &  3.762\errr{160}{127} & [ 7,11] & 17.48 / 4  \\ \hline 
\hline
\end{tabular}
\end{center}
\label{TAB_nucleon_V16X24}
\caption{Nucleon masses on $16^3\cdot24$, in lattice units and in
units of $r_0$.} 
\end{table}

\begin{table}
\begin{center}
\begin{tabular}{|ccccccc|}\hline\hline
$L^3\cdot T$ & $\kappa^{sea}$ & $\kappa^{val}$ & $m_{\Delta}a$ & $m_{\Delta}r_0$ & Fit & $\chi^2 /$ dof  \\ \hline \hline
$8^3\cdot 24$ & 0.1370 & 0.1370 &  2.146\err{ 39}{ 13} &  4.799\errr{121}{103} & [ 7,11] &  5.73 / 4  \\
              &  & 0.1380 &  2.070\err{ 40}{ 14} &  4.629\errr{120}{100} & [ 7,11] &  4.79 / 4  \\
              &  & 0.1390 &  1.989\err{ 44}{ 13} &  4.448\errr{125}{ 96} & [ 7,11] &  4.64 / 4  \\
              &  & 0.1395 &  1.947\err{ 46}{ 12} &  4.353\errr{128}{ 94} & [ 7,11] &  4.75 / 4  \\ \hline 
$8^3\cdot 24$ & 0.1380 & 0.1370 &  2.131\err{ 51}{  1} &  5.274\errr{202}{143} & [ 6,11] & 16.72 / 6  \\
              &  & 0.1380 &  1.918\err{ 37}{  4} &  4.747\errr{169}{129} & [ 7,11] &  6.51 / 4  \\
              &  & 0.1390 &  1.854\err{ 34}{  8} &  4.588\errr{160}{126} & [ 7,11] &  3.99 / 4  \\
              &  & 0.1395 &  1.831\err{ 33}{ 12} &  4.531\errr{158}{126} & [ 7,11] &  3.93 / 4  \\ \hline 
$8^3\cdot 24$ & 0.1390 & 0.1370 &  1.790\err{ 34}{  9} &  5.174\errr{145}{107} & [ 6,11] & 17.07 / 6  \\
              &  & 0.1380 &  1.717\err{ 32}{  8} &  4.964\errr{139}{102} & [ 6,11] & 14.01 / 6  \\
              &  & 0.1390 &  1.651\err{ 27}{  9} &  4.773\errr{127}{100} & [ 6,11] & 11.26 / 6  \\
              &  & 0.1395 &  1.621\err{ 27}{ 13} &  4.687\errr{124}{101} & [ 6,11] & 10.62 / 6  \\ \hline 
$8^3\cdot 24$ & 0.1395 & 0.1370 &  1.718\err{ 42}{  6} &  6.389\errr{219}{126} & [ 6,11] & 10.07 / 6  \\
              &  & 0.1380 &  1.644\err{ 42}{  6} &  6.111\errr{214}{121} & [ 6,11] &  9.26 / 6  \\
              &  & 0.1390 &  1.578\err{ 38}{ 13} &  5.866\errr{200}{124} & [ 6,11] & 10.38 / 6  \\
              &  & 0.1395 &  1.538\err{ 43}{ 19} &  5.718\errr{210}{130} & [ 7,11] &  7.59 / 4  \\ \hline 
\hline
\end{tabular}
\end{center}
\label{TAB_delta_V8X24}
\caption{$\Delta$ masses on $8^3\cdot24$, in lattice units and in
units of $r_0$.} 
\end{table}

\begin{table}
\begin{center}
\begin{tabular}{|ccccccc|}\hline\hline
$L^3\cdot T$ & $\kappa^{sea}$ & $\kappa^{val}$ & $m_{\Delta}a$ & $m_{\Delta}r_0$ & Fit & $\chi^2 /$ dof  \\ \hline \hline
$12^3\cdot 24$ & 0.1370 & 0.1370 &  2.139\err{ 29}{  1} &  4.907\errr{ 79}{ 49} & [ 7,11] & 10.07 / 4  \\
               &  & 0.1380 &  2.071\err{ 27}{  3} &  4.750\errr{ 75}{ 48} & [ 7,11] & 10.48 / 4  \\
               &  & 0.1390 &  2.000\err{ 25}{  4} &  4.587\errr{ 70}{ 47} & [ 7,11] & 10.97 / 4  \\
               &  & 0.1395 &  1.966\err{ 23}{  5} &  4.511\errr{ 65}{ 47} & [ 7,11] & 11.01 / 4  \\ \hline 
$12^3\cdot 24$ & 0.1380 & 0.1370 &  1.933\err{ 30}{  1} &  4.963\errr{101}{ 58} & [ 7,11] &  8.07 / 4  \\
               &  & 0.1380 &  1.855\err{ 30}{  2} &  4.763\errr{ 99}{ 56} & [ 7,11] &  6.26 / 4  \\
               &  & 0.1390 &  1.773\err{ 32}{  2} &  4.553\errr{102}{ 53} & [ 7,11] &  5.89 / 4  \\
               &  & 0.1395 &  1.754\err{ 22}{ 14} &  4.504\errr{ 82}{ 64} & [ 7,11] &  5.19 / 4  \\ \hline 
$12^3\cdot 24$ & 0.1390 & 0.1370 &  1.714\err{ 31}{  5} &  5.221\errr{136}{ 90} & [ 7,11] & 18.87 / 4  \\
               &  & 0.1380 &  1.630\err{ 30}{  6} &  4.964\errr{131}{ 87} & [ 7,11] & 16.59 / 4  \\
               &  & 0.1390 &  1.545\err{ 27}{  6} &  4.707\errr{123}{ 83} & [ 7,11] & 13.88 / 4  \\
               &  & 0.1395 &  1.502\err{ 29}{  6} &  4.576\errr{124}{ 80} & [ 7,11] & 12.39 / 4  \\ \hline 
$12^3\cdot 24$ & 0.1395 & 0.1370 &  1.539\err{ 26}{ 10} &  5.285\errr{115}{ 80} & [ 7,11] & 26.22 / 4  \\
               &  & 0.1380 &  1.457\err{ 20}{ 16} &  5.005\errr{ 97}{ 87} & [ 7,11] & 21.69 / 4  \\
               &  & 0.1390 &  1.372\err{ 19}{ 19} &  4.711\errr{ 92}{ 91} & [ 7,11] & 17.49 / 4  \\
               &  & 0.1395 &  1.329\err{ 22}{ 20} &  4.565\errr{ 99}{ 93} & [ 7,11] & 15.43 / 4  \\ \hline 
$12^3\cdot 24$ & 0.1398 & 0.1380 &  1.373\err{ 34}{ 21} &  5.016\errr{129}{ 83} & [ 7,11] & 34.34 / 4  \\
               &  & 0.1390 &  1.273\err{ 32}{ 26} &  4.648\errr{121}{ 99} & [ 7,11] & 27.18 / 4  \\
               &  & 0.1395 &  1.222\err{ 32}{ 27} &  4.463\errr{121}{103} & [ 7,11] & 23.00 / 4  \\
               &  & 0.1398 &  1.191\err{ 29}{ 28} &  4.349\errr{113}{105} & [ 7,11] & 20.55 / 4  \\ \hline 
\hline
\end{tabular}
\end{center}
\label{TAB_delta_V12X24}
\caption{$\Delta$ masses on $12^3\cdot24$, in lattice units and in
units of $r_0$.} 
\end{table}

\begin{table}
\begin{center}
\begin{tabular}{|ccccccc|}\hline\hline
$L^3\cdot T$ & $\kappa^{sea}$ & $\kappa^{val}$ & $m_{\Delta}a$ & $m_{\Delta}r_0$ & Fit & $\chi^2 /$ dof  \\ \hline \hline
$16^3\cdot 24$ & 0.1390 & 0.1390 &  1.599\err{ 34}{ 20} &  4.838\errr{116}{ 73} & [ 6,11] & 32.43 / 6  \\
               &  & 0.1395 &  1.555\err{ 34}{ 19} &  4.704\errr{114}{ 68} & [ 6,11] & 28.16 / 6  \\
               &  & 0.1398 &  1.532\err{ 35}{ 19} &  4.635\errr{115}{ 69} & [ 6,11] & 25.28 / 6  \\ \hline 
$16^3\cdot 24$ & 0.1395 & 0.1390 &  1.395\err{ 28}{  5} &  4.805\errr{111}{ 81} & [ 7,11] & 11.43 / 4  \\
               &  & 0.1395 &  1.347\err{ 30}{  6} &  4.640\errr{115}{ 79} & [ 7,11] & 10.63 / 4  \\
               &  & 0.1398 &  1.318\err{ 30}{  8} &  4.538\errr{116}{ 80} & [ 7,11] & 10.20 / 4  \\ \hline 
$16^3\cdot 24$ & 0.1398 & 0.1390 &  1.296\err{ 29}{ 11} &  4.733\errr{112}{ 55} & [ 7,11] & 19.39 / 4  \\
               &  & 0.1395 &  1.244\err{ 27}{ 11} &  4.541\errr{106}{ 56} & [ 7,11] & 17.22 / 4  \\
               &  & 0.1398 &  1.208\err{ 29}{ 12} &  4.411\errr{111}{ 56} & [ 7,11] & 15.52 / 4  \\ \hline 
\hline
\end{tabular}
\end{center}
\label{TAB_delta_V16X24}
\caption{$\Delta$ masses on $16^3\cdot24$, in lattice units and in
units of $r_0$.} 
\end{table}

\end{document}